\documentclass[conference]{IEEEtran}
\IEEEoverridecommandlockouts
\usepackage{booktabs}
\usepackage{bm}
\usepackage{pifont}
\usepackage{subcaption} 
\usepackage{cite}
\usepackage{amsmath,amssymb,amsfonts}
\usepackage{algorithmic}
\usepackage{graphicx}
\usepackage{textcomp}
\usepackage{xcolor}
\def\BibTeX{{\rm B\kern-.05em{\sc i\kern-.025em b}\kern-.08em
    T\kern-.1667em\lower.7ex\hbox{E}\kern-.125emX}}

\begin{document}

\title{Rank-Based Modeling for Universal Packets Compression in Multi-Modal Communications}

\author{
		\IEEEauthorblockN{Xuanhao Luo,
        Zhiyuan Peng,
        Zhouyu Li,
        Ruozhou Yu,
        Yuchen Liu
 }
		\IEEEauthorblockA{
		North Carolina State University, USA
		}
	}
\maketitle

\begin{abstract}

The rapid increase in networked systems and data transmission requires advanced data compression solutions to optimize bandwidth utilization and enhance network performance. This study introduces a novel byte-level predictive model using Transformer architecture, capable of handling the redundancy and diversity of data types in network traffic as byte sequences. Unlike traditional methods that require separate compressors for different data types, this unified approach sets new benchmarks and simplifies predictive modeling across various data modalities such as video, audio, images, and text, by processing them at the byte level. This is achieved by predicting subsequent byte probability distributions, encoding them into a sparse rank sequence using lossless entropy coding, and significantly reducing both data size and entropy. 
Experimental results\textsuperscript{1} show that our model achieves compression ratios below 50\%, while offering models of various sizes tailored for different communication devices. Additionally, we successfully deploy these models on a range of edge devices and servers, demonstrating their practical applicability and effectiveness in real-world network scenarios. This approach significantly enhances data throughput and reduces bandwidth demands, making it particularly valuable in resource-constrained environments like the Internet of Things sensor networks. 
\footnotetext[1]{{The source code of our proposed ByteTrans model is available at \textit{https://github.com/Xuanhao-Luo/ByteTrans}.}}

\end{abstract}


\section{Introduction}
\noindent
As networked systems continue to evolve, the interconnection and intelligent management of devices through the Internet of Things (IoT) become increasingly vital. The IoT integrates a wide range of network devices with standardized communication protocols, enabling sophisticated interaction and automation across various applications \cite{li2015internet, nguyen20216g}. This advancement necessitates efficient data transmission strategies to handle the increased data flow while optimizing network resource utilization. In the context of IoT, multi-modal communications are essential because they allow a more comprehensive and flexible interaction between devices, users, and systems by enabling the integration of various forms of data from different sources.
However, modern communication backbone such as wireless local area networks (WLANs) and sensor networks faces the dual challenges of limited bandwidth \cite{sinha2017survey, hsu2017breaking} and high data redundancy \cite{kumar2018strategy, verma2018data}, making efficient data management and communication even more critical for optimal performance.
Packet data, for instance, which includes IP addresses and other header information, often exhibits significant redundancy. These elements are highly repetitive across numerous packets, indicative of the structured nature of entangled communications and presenting a prime opportunity for optimization.
Efficient packet compression strategies can significantly alleviate bandwidth constraints by reducing the data load transmitted across the network, thereby enhancing the overall performance and scalability of multi-modal communications and their applications.

Building upon Claude Shannon's foundational work \cite{shannon1951prediction}, which explored the connection between \textit{prediction} and \textit{compression} to estimate the entropy of the English language, this research delves into enhancing data compression efficacy through advanced byte-level predictive modeling. Shannon posited that effective predictors of sequential data values could be directly leveraged to improve compression algorithms, which typically involve two stages: modeling and coding \cite{cox2016syntactically}. The coding stage, refined to near-optimal levels, employs entropy coding techniques, a lossless data compression method that utilizes variable-length encoding to assign shorter codes to more frequently occurring symbols. This approach ensures efficiency close to the theoretical limit of \(\log_2 \frac{1}{p}\) bits per symbol, where \(p\) is the probability of a symbol's occurrence. 
However, the modeling stage, which involves estimating the likelihood of subsequent symbols based on historical data, remains a complex and less explored challenge. The overall compression ratio is predominantly determined by the accuracy of this predictive model. Our work herein seeks to advance the state of the art in data predictive modeling,  thereby transforming redundant multi-modal communication packets into a sparse format, which facilitates efficient compression using entropy coding methods.

The primary challenge in enhancing compression algorithms for networked systems lies in the accurate prediction of data contained within the packets, as network traffic always comprises a diverse array of data modalities, including video, audio, images, and text \cite{tian2024synchronous}, each with \textit{distinct} characteristics and encoding schemes. Traditionally, handling such heterogeneity in data requires developing specialized encoding models for each data type, such as High Efficiency Video Coding (HEVC) for video \cite{sullivan2012overview} and JPEG for image compression \cite{marcellin2000overview}.
However, when viewed from the perspective of a communication network, regardless of the data types, all information is ultimately encoded and transmitted as a sequence of bytes. This ubiquitous byte-level representation presents an opportunity to streamline the predictive modeling process across all data modalities.
In this context, our research introduces a novel byte-level mechanism, \textit{ByteTrans}, that capitalizes on the fundamental binary format of network data. By focusing on bytes as the basic unit of data, our model bypasses the need for multiple modality-specific predictors, leading to a universal packet compression paradigm. 
This one-size-fits-all approach simplifies predictive modeling for data compression while improving communication efficiency, using a Transformer-based framework to predict the next byte by leveraging redundancy and patterns in byte-encoded data streams.
In summary, this work presents the following contributions:
\begin{itemize}
    \item To the best of our knowledge, this is the first study to employ a predictive model for compressing network-level packets within multi-modal communications using a byte-level Transformer as predictor. This facilitates the handling of diverse data modalities within a unified framework, enhancing the compression process through a novel \textit{rank-based} encoding scheme.

    \item We develop several ByteTrans variants with different model sizes, tailored to suit various network conditions and deployments. This adaptive design ensures versatility, allowing the approach to be optimized for a range of computational capacities, from high-performance servers to resource-constrained devices.

    \item {The experimental results reveal that the entropy of elaborately compressed ranks is reduced to less than 50\% of the original data packet entropy, achieving a compression ratio of 46.3\%. These findings highlight the models' effectiveness in significantly reducing data redundancy. 
    Notably, our predictive framework achieves up to a 14.6\% lower compression ratio compared to the baseline data compressor, demonstrating its enhanced capability in optimizing data transmission.}

    \item Our models are rigorously tested across a spectrum of network settings and devices, affirming their adaptability and efficiency. We conduct real-world deployments and configurations of these models on various edge devices and servers. These validations confirmed the ByteTrans operational viability and demonstrated its robust performance, reinforcing the potential for broad application in diverse computational and network environments.

\end{itemize}

\section{Related Works}
\noindent
The foundational work by Shannon in \cite{shannon1951prediction} established that effectively predicting the next element in a sequence is crucial for developing an efficient compression algorithm. Building upon this concept, various studies have explored the use of deep neural networks for data compression. \cite{schmidhuber1996sequential} employed Recurrent Neural Networks (RNNs) as predictive models for lossless text compression, while \cite{mahoney2000fast} introduced a two-layer maximum entropy model combined with a predictive arithmetic encoder. 
In \cite{cox2016syntactically}, the author introduce a semantic compression approach using RNNs and syntactic information, focusing on leveraging grammatical structures to improve compression efficiency rather than traditional lossless methods.
\cite{goyal2018deepzip} introduces DeepZip, an RNN-based framework for lossless compression of sequential data, such as text and genomic sequences, to predict the likelihood of the next symbol based on prior context.
More recently, \cite{valmeekam2023llmzip} introduces LLMZip, a novel approach for lossless text compression using large language models (LLMs) like LLaMA-7B as predictors for the next token based on past tokens. This method estimates a new asymptotic upper bound on the entropy of text, achieving a significantly lower entropy compared to existing estimates.
Despite these advancements, none have specifically addressed the challenges of compressing data packets in communication networks, such as in the form of multi-modal streams and efficient transmissions. Text-focused approaches struggle to apply here because they are not designed to handle the diverse data types and time-series constraints typical of network traffic.

Following the exploration of neural network applications in compression, the role of sequence data prediction has gained prominence, traditionally addressed using models like RNNs \cite{sutskever2011generating}, Long Short-Term Memory (LSTM) \cite{10.1162/neco.1997.9.8.1735}, and Gated recurrent units (GRU) \cite{cho2020learning}. These models sequentially process time-series data, maintaining a hidden state to capture temporal dependencies, but they struggle with long sequences due to vanishing or exploding gradient issues \cite{pascanu2013difficulty}. The introduction of the Transformer architecture \cite{vaswani2017attention} revolutionized this task by using self-attention mechanisms to model dependencies across entire sequences simultaneously, enabling more effective handling of long-range dependencies \cite{wu2020deep}, \cite{wu2020adversarial}. 
Recently, LLMs increasingly use decoder-only Transformer architectures due to their optimal design for generative tasks, computational efficiency, flexibility across various applications, and strong parallelization capabilities. Models like Llama3 \cite{dubey2024llama}, GPT-4 \cite{achiam2023gpt}, and Google Gemini \cite{anil2023gemini} exemplify this trend by excelling in autoregressive text generation, enabling them to handle a wide range of natural language processing tasks within a unified framework while minimizing computational and memory overhead.

However, these Transformer-based LLMs are typically token-based, which poses challenges when dealing with data modalities such as network packets that do not conform easily to tokenization, as they are always structured as continuous byte streams. 
To address this, \cite{yu2023megabyte} introduces a multiscale autoregressive transformer model designed to efficiently handle million-byte sequences without tokenization, leveraging both global and local context modeling to improve scalability and performance across various modalities, including text, image, and audio. 
Similarly, other works such as 
\cite{wang2024mambabyte, perez2024compressed, wu2024beyond, han2024jpeg} also demonstrate that byte-based compressed-language models can effectively understand, handle, and generate 
different data types directly from their compressed byte streams. 
Despite these advances, a gap remains in the application of byte-based models to process network-level data packets, which exhibit specific underlying patterns, such as packet headers, inter-packet dependencies, and strict timing constraints, that those compressed-language models cannot effectively handle due to their focus on language structure. Furthermore, their overly general predictive capabilities may become a burden when deployed in real communication networks, where computational efficiency and lightweight deployment are critical, especially in IoT applications.

To overcome these limitations, as our work focuses here, it is critical to address the unique challenge of modeling network traffic packets, which encompass diverse data modalities but can essentially be represented as continuous byte streams using a well-designed byte-based Transformer decoder. By marrying this predictive model with lossless compression schemes, the data packets can be effectively compressed to a new level, leveraging the inherent redundancy within packet sequences for substantial size reduction.

\section{ByteTrans Model Architecture}
\noindent
This section introduces the proposed model architecture, focusing on the Transformer's ability to predict and compress byte sequences efficiently. We begin with a theoretical overview, explaining how entropy and information theory underpin our approach to data compression. We then utilize the Transformer architecture, specifically its decoder component, for high-accuracy probability estimation of byte sequences. This leads to precise \textit{ranking} of bytes according to their likelihood, which  significantly reduces data size when encoded using entropy coding techniques. Subsequently, we discuss entropy-based lossless compression methods and adapt model sizing to match various resource-constraint environments.

\subsection{ByteTrans at a Glance}
\noindent
\textbf{Rationale.} At the heart of information theory is the concepts of entropy, which provide a theoretical foundation for data compression techniques. Entropy, denoted as \(H(X)\), is defined for a data stream \(X\) and quantifies the average uncertainty inherent in the sequence as:
\begin{equation}
H(X) = -\sum_{i} p(x_i) \log_2 p(x_i),
\end{equation}
where \(p(x_i)\) is the probability of occurrence of symbol \(x_i\). The entropy not only encapsulates the amount of information contained within a sequence but also establishes a lower bound on the number of bits required to encode the sequence without loss. Correspondingly, the optimal coding length for a symbol \(x_i\) becomes:
\begin{equation}
L(x_i) = \log_2 \frac{1}{p(x_i)}.
\end{equation}
This suggests that any symbols occurring more frequently, i.e. with higher \(p(x_i)\), can be encoded using fewer bits, thereby reducing the average length of the encoded data.

Building upon these principles, we aim to design a data compression system that concentrates the probability distributions of symbols as much as possible. By increasing the predictability of symbols--ensuring that certain symbols are far more likely than others, thereby boosting \(p(x_i)\) for these symbols--we can improve the compression ratio to a new level. This can be achieved by exploiting a Transformer-based model, specifically adapting the decoder component, which is naturally suited for processing sequential data without relying on the temporal dependencies of recurrent models.

\noindent \textbf{Rank Transformation.} Specifically, the core functionality of the Transformer decoder is to estimate the probability distribution over a byte sequence \( \mathcal{S}_N = \{S_1, S_2, \ldots, S_N\} \), where each \( S_i \) represents a byte. Then, the Transformer can predict the probability distribution \( P(S_{i+1} | S_1, \ldots, S_i) \) for the next byte \( S_{i+1} \) based on its predecessors as:
\begin{equation}
P(S_{i+1} | S_1, \ldots, S_i) = {\rm softmax}\{f_{Decoder}(S_1, \ldots, S_i)\}.
\end{equation}
Following the probability estimation, ByteTrans ranks all possible subsequent bytes based on their likelihoods. The rank of the $i-th$ byte $r_i$ can be represented as:
\begin{equation}
r_i = \text{\textbf{Rank} of } S_i \text{ based on } P(S_i | S_1, \ldots, S_{i-1}),
\end{equation}
Then, the sequence of bytes \( \mathcal{S} = \{S_1, S_2, \ldots, S_N\} \) are transformed into a sequence of ranks \(r = \{r_1, r_2, \ldots, r_N\} \). 

\begin{figure}[t]
	\centerline{\includegraphics[scale=0.51]{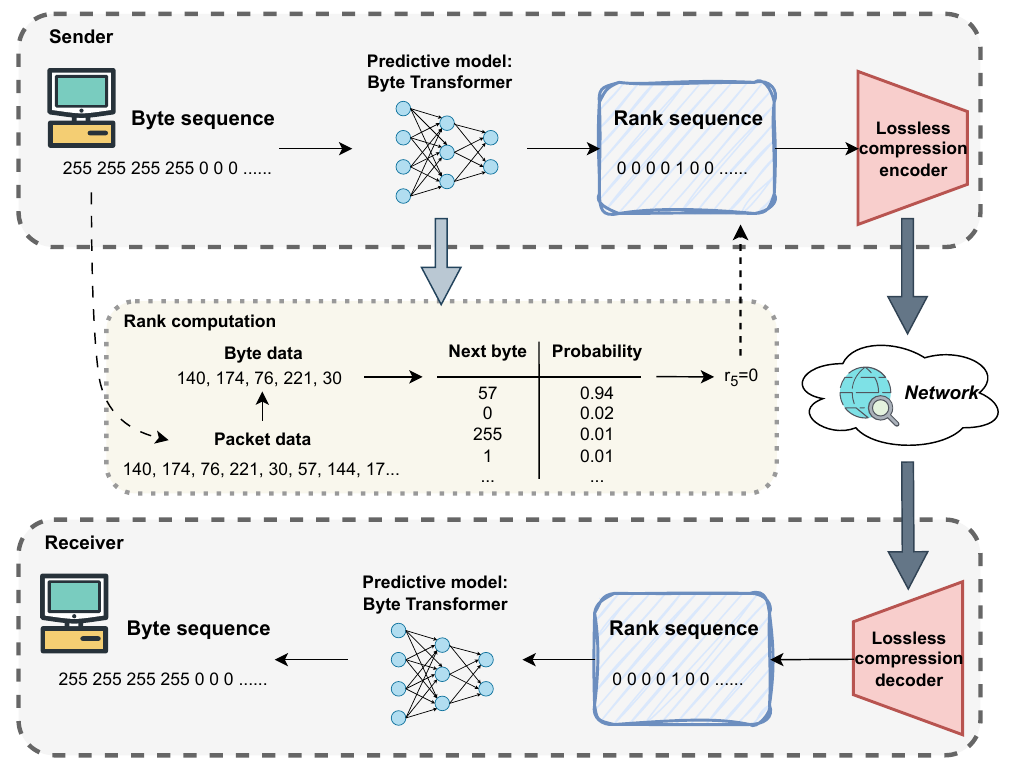}}
    \caption{The overview of ByteTrans framework.} 
	\label{framework}
    \vspace{-5mm}
\end{figure}

\noindent \textbf{Lossless Compression.} Next, the lossless compression techniques can be employed to encode these ranks, rather than the original bytes, into a compact binary format. Based on the Transformer's precision, the resulting rank sequences exhibit substantial redundancy, with frequent and predictable patterns of low ranks such as 0 or 1. By leveraging the entropy coding methods like Huffman coding or arithmetic coding, these redundant and predictable patterns can be substantially compressed, thereby reducing the overall size of the encoded data. In essence, these techniques can adjust the coding length $L(x_i)$ based on the frequency of occurrence, assigning shorter codes to more frequent ranks. This optimizes the compression ratio by efficiently utilizing the statistical dependencies uncovered by the Transformer.

\noindent \textbf{Reversed Decoding.} Once the receiver obtains the message through communication channels, it will decode it with the same coding method and predictive model. The decompression process is designed to be symmetrical but begins with an additional initial step. First, the decompression algorithm reconstructs the sequence of ranks from the compressed data.
Once the ranks $r_i$ are restored, the same Transformer model used during compression is employed to predict the probability distributions for each byte position. By aligning these predicted distributions with the recovered ranks, the original byte sequence can be effectively reconstructed. This process ensures that the data integrity and fidelity are preserved throughout both the compression and decompression cycles. The overall design process is shown in Figure~\ref{framework}, with each block in the model architecture detailed as follows.


\subsection{Byte Transformer for Probability Prediction}
\noindent
The core of ByteTrans lies in its rank-based transformation powered by a Transformer architecture, fundamentally reshaping the way sequential data is processed. While the complete architecture consists of both encoder and decoder components, for the purpose of byte-level data compression, this model block exclusively utilizes the Transformer decoder, which is adept at generating sequences based on learned relationships from data. The Transformer decoder is composed of a stack of identical layers, each containing two main sub-modules, i.e., a multi-head self-attention mechanism and a position-wise fully connected feed-forward network.

\begin{figure}[t]
\centerline{\includegraphics[scale=0.6]{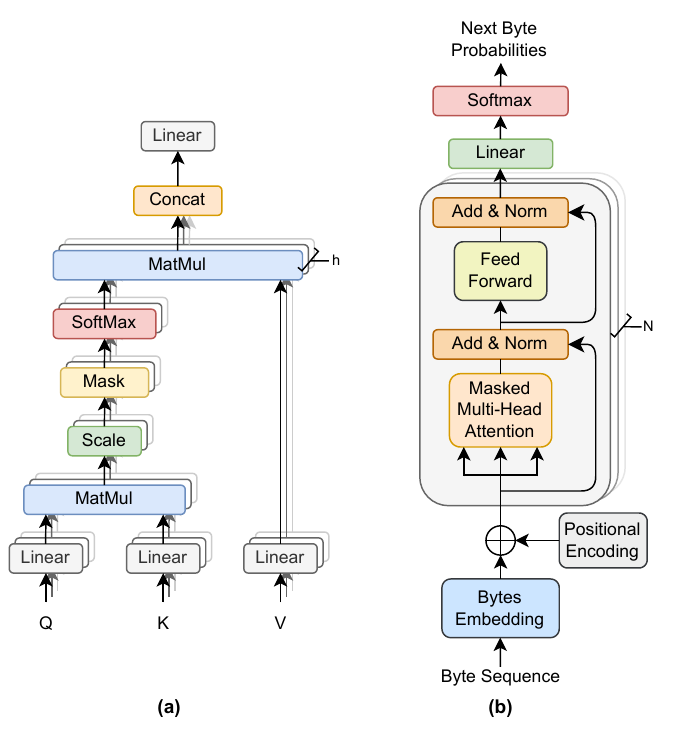}}
     \caption{(a) Architecture of multi-head attention. (b) Architecture of byte Transformer decoder.} 
	\label{Transformer}
    \vspace{-6mm}
\end{figure}

Here, the self-attention mechanism is used to weigh the influence of different parts of the input data independently of their positions in the sequence. Given a sequence of input embedding $X$, the self-attention scores are calculated using three sets of weights, namely Query ($Q$), Key ($K$), and Value ($V$) matrices, which are learned during training:
\begin{equation}
    Q = XW^Q, \quad K = XW^K, \quad V = XW^V.
\end{equation}
The attention scores between each pair of positions are computed by the dot product of the $Q$ and the transposed Key matrix, followed by a scaling factor and a softmax to obtain the weights:
\begin{equation}
    Attention(Q, K, V) = softmax\left(\frac{QK^T}{\sqrt{d_k}}\right)V,
\end{equation}
where $d_k$ is the dimension of the key vectors which scales the dot product to prevent overly large values. To ensure that predictions for a given position depend only on known outputs from preceding positions—crucial for our data compression task, where future data streams should not influence past predictions—masking is applied within the self-attention calculations. This masking is applied by adding a mask matrix to the attention scores before applying softmax, where the mask entries are set to negative infinity for positions that should not influence predictions, effectively nullifying their impact.
To allow the model to jointly attend to information from different representation subspaces at different positions, the self-attention mechanism is extended to multi-head attention:
\begin{equation}
    MultiHead(Q, K, V) = Concat(H_1, \dots, H_h)W^O,
\end{equation}
where each head is defined as:
\begin{equation}
    H_i = Attention(QW_i^Q, KW_i^K, VW_i^V).
\end{equation}
This allows the model to capture multiple dependencies, such as syntactic and semantic, simultaneously. The detailed architecture of multi-head attention mechanism is shown in Figure~\ref{Transformer} (a).

Particularly, each layer of the Transformer's decoder also includes a position-wise fully connected feed-forward network ($FFN$), which is applied identically to each position:
\begin{equation}
    FFN(x) = \text{ReLU}(xW_1 + b_1)W_2 + b_2,
\end{equation}
where \(x\) represents the input to the feed-forward network. \(W_1\) and \(b_1\) are the weights and biases of the first linear transformation, while \(W_2\) and \(b_2\) are those of the second linear transformation.

Since the self-attention mechanism does not inherently consider the data sequence order, positional encodings are added to the input embeddings to provide this information. These encodings, defined using sinusoidal functions, represent the position of each element in the sequence and are expressed as:
\begin{equation}
\begin{gathered}
PE_{(pos, 2i)} = \sin\left(\frac{pos}{10000^{2i/d_{\text{m}}}}\right), \\
PE_{(pos, 2i+1)} = \cos\left(\frac{pos}{10000^{2i/d_{\text{m}}}}\right),
\end{gathered}
\end{equation}
where $pos$ is the position index of a byte in the input sequence, $i$ is the index of the specific dimension in the positional encoding vector being calculated, and $d_{\text{m}}$ is the embedding dimension.

To facilitate training deep learning models, each sub-layer in the transformer decoder has a residual connection around it followed by layer normalization:
\begin{equation}
    LayerNorm(x + Sublayer(x)),
\end{equation}
where $Sublayer(x)$ is the function implemented by the attention or the feed-forward network. 
The output from the decoder's top layer is transformed into a probability distribution over all possible next bytes using a linear layer followed by a softmax function:
\begin{equation}
    Output= softmax(zW + b),
\end{equation}
where $z$ is the output from the last network layer, and $W$ and $b$ are the weights and bias of the output layer. The overall model of byte Transformer is shown in Figure~\ref{Transformer} (b).

During the training phase, we utilize the cross-entropy loss to optimize the prediction of the next byte in the data sequence. The cross-entropy loss measures the difference between the predicted probability distribution and the actual distribution, where the actual distribution is typically represented as a one-hot encoded vector of the true next byte in the sequence. In the context of byte prediction, each class represents one of the possible byte values, which ranges from 0 to 255, resulting in the total of 256 classes. As such, the loss is calculated as follow:
\begin{equation}
    \mathcal{L} = -\sum_{c=0}^{255} y_{o,c} \log(p_{o,c}),
\end{equation}
where \(y_{o,c}\) is a one-hot encoded vector, with the element corresponding to the actual byte value \(c\) for observation \(o\) set to 1, and all other elements set to 0. The term \(p_{o,c}\) represents the predicted probability that the byte value for observation \(o\) is \(c\). 
By minimizing this cross-entropy loss, the model effectively increases the likelihood of predicting the correct subsequent byte based on the given context.
Enhancing the predictive accuracy ensures that the ranks generated for the following compression process are as informative and compact as possible, thereby maximizing the compression ratio while maintaining lossless reconstruction capabilities.


\subsection{Lossless Compression Methodology}
\noindent
After the byte Transformer predicts the probability distribution for the next byte in a sequence and converts it into a sequence of ranks, the next step involves compressing these ranks using lossless entropy coding techniques. Essentially, these methods exploit the statistical redundancy in our transformed rank sequence by assigning shorter codes to more frequent ranks. This approach significantly reduces the size of the transmitted data, as frequent ranks can be always encoded with fewer bits.
For the actual data compression, we employ the zlib compression library \cite{zlib}, which is renowned for its effectiveness and efficiency in compressing data streams. Zlib utilizes a combination of the Deflate compression algorithm, which employs a mixture of the LZ77 \cite{ziv1977universal} algorithm and Huffman coding \cite{huffman1952method}, providing robust and adjustable compression capabilities. This method adapts dynamically to the data characteristics, ensuring optimal compression ratios and performance, particularly beneficial in environments where bandwidth is a constraint.

To quantitatively assess the effectiveness of the compression method, we utilize the compression ratio \(\rho\), calculated as:
\[
\rho = \frac{S_c}{S_o},
\]
where $S_c$ refers to the size of the data after applying the compression algorithm, and $S_o$ refers to the size of the data before compression. A lower \(\rho\) indicates that the compressed data size is significantly smaller compared to the original size, implying a more effective compression algorithm. This metric serves as the primary measure of performance, guiding the evaluation of different models and compression techniques to identify the most efficient approach for reducing data sizes without losing information.

\subsection{Model Adaptation}
\noindent
In modern networked environments, spanning from robust data center servers \cite{ma2023intelligent} to resource-constrained IoT devices \cite{li2015internet}, and extending to the highly dynamic and interconnected edge systems and Internet of Vehicles \cite{luo2023clothoid}, the disparity in available computational resources presents significant challenges. To address the computationally intensive nature and high inference latency of transformer-based models, we implement a model adaptation strategy, involving pretraining models of varying sizes to dynamically adjust based on the available computational resources, aligning with user requirements and specific network conditions. This adaptive approach ensures seamless integration and operational efficiency across a wide range of IoT platforms.

By pretraining models of varying sizes, our adaptation strategy optimizes resource efficiency and ensures deployment-specific configurations. Moreover, varying link quality in wireless networks \cite{luo2024rm, luo2025denoising} calls for adaptive compression models for optimal efficiency. Larger models, capable of capturing complex data patterns due to their enhanced computational power, are ideal for high-capacity servers where they can operate without constraints. Conversely, smaller models are particularly effective in environments with repetitive data patterns, such as WLANs or IoT setups, where data redundancy is substantial. These smaller models quickly learn and compress common patterns, significantly reducing both training and inference costs, making them well-suited for environments with limited resources.
Our comprehensive evaluation, detailed in Section \ref{evaluation}, assesses the performance of these adapted models across various real-world devices, highlighting their efficacy and scalability.

\section{Performance Evaluation}
\label{evaluation}
\noindent
In this section, we present a comprehensive evaluation of the performance of our proposed universal data compression models. We begin by detailing the dataset and preprocessing steps, including the preparation of network packets for input into ByteTrans. Following this, we analyze the performance of ByteTrans with different model sizes in terms of their ability to predict byte sequences. Lastly, we assess the efficiency of these predictive models across various devices, examining key metrics such as compression time, CPU/GPU/RAM usage, and power consumption, showcasing their adaptability in diverse communication environments.

\subsection{Dataset and Preprocessing}
\noindent
In our experiments, we process data from the dataset \cite{dadkhah2024ciciomt2024} collected in a controlled lab environment, specifically focusing on the benign wireless data subset. This dataset is selected due to its comprehensive coverage of various IoT devices operating within a real-world healthcare environment, capturing a wide range of normal operational behaviors. It is meticulously designed to emulate realistic Internet of Medical Things (IoMT) network traffic, providing a valuable resource for evaluating network traffic and protocols. From this dataset, we extract around 31,000 data packets of benign Wi-Fi traffic. These packets are chosen to represent typical, non-malicious activity within a healthcare IoT network, serving as a baseline for comparison in our analysis. For testing purposes, we select 2,560 packets from the dataset., while the majority of experiments utilizes 18,192 packets for training. Additionally, we explore the impact of training data volume by using a smaller training dataset of 8,960 packets and a larger version with 28,192 packets. Each packet contains detailed byte-level data information, encompassing various protocol headers and payload information, sourced from a range of IoMT devices such as medical monitors (time-series data), smart cameras (image data), and environmental sensors (structured texts), all communicating over standard Wi-Fi protocols. This multi-modal data collection ensures a comprehensive 
representation of typical IoT operations, spanning various data types and communication patterns essential for robust model training and evaluation.

\begin{figure}[ht]
\vspace{-2mm}
	\centerline{\includegraphics[scale=0.23]{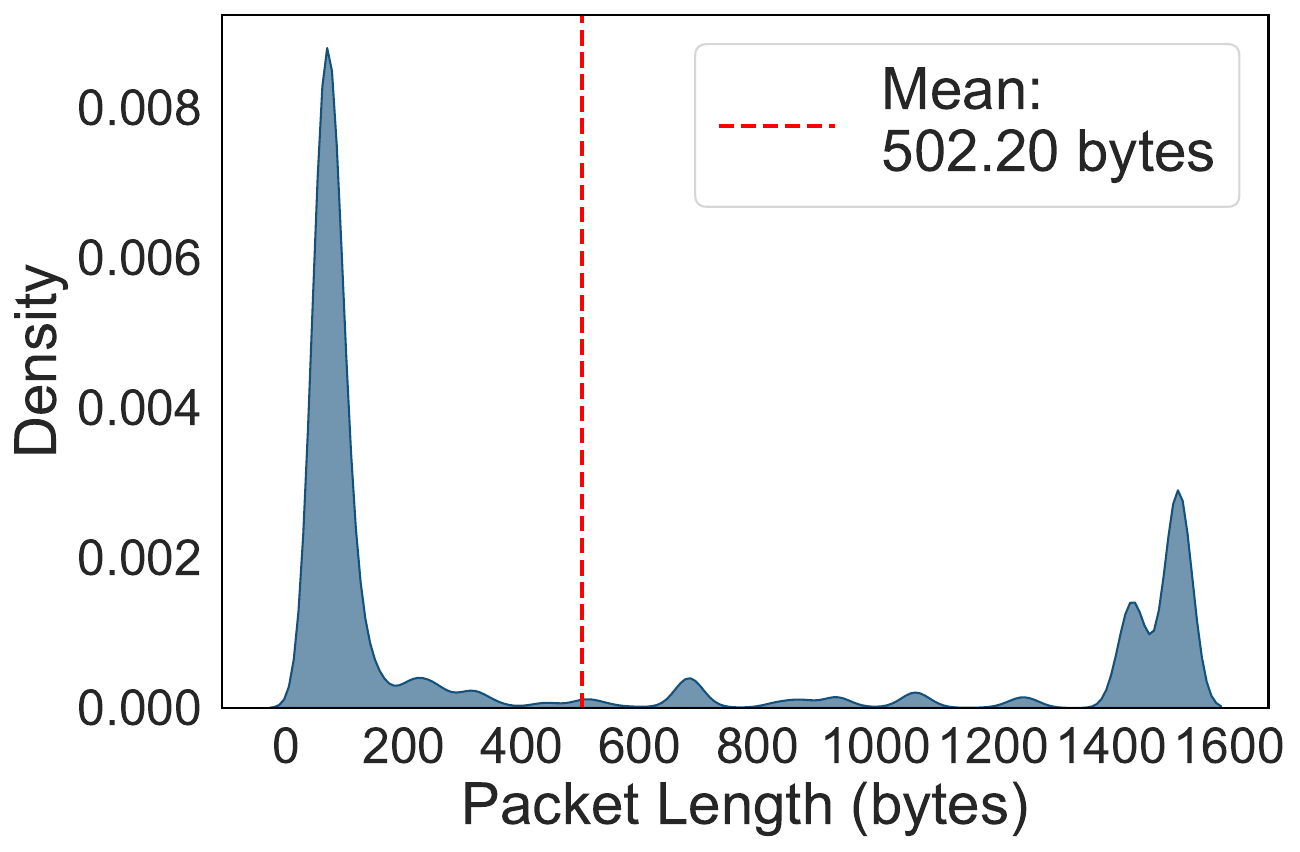}}
     \caption{Packets length distribution (KDE).} 
	\label{distribution}
    \vspace{-3mm}
\end{figure}

Figure \ref{distribution} illustrates the packet length distribution of the dataset, visualized using a Kernel Density Estimation (KDE) plot, which provides a smooth estimation of the probability density function of packet lengths. 
This graphical analysis indicates a mean packet length of 502.20 bytes, with most packet lengths are concentrated below 200 bytes. This clustering of packet sizes reflects the typical pattern of IoT communications, where data transmissions are often brief and frequent, rather than large data transfers.
Given these observations, the sequence length for data preprocessing is set at 256 bytes, capturing the majority of packets in their entirety without requiring fragmentation.

To prepare the data for our predictive model training, we start by converting raw packet data into a trainable format. First, we load the binary data and use a distinct byte sequence, \texttt{b'\textbackslash x00\textbackslash xFF\textbackslash x00\textbackslash xFF'}, as a separator to delineate packets, ensuring clear packet boundaries since this sequence does not appear in the actual data. Each packet is then transformed into a sequence of integers, converting the bytes from \texttt{uint8} to \texttt{int16} to accommodate special markers. We append an end-of-packet marker (256) and use padding (257) to standardize sequence lengths, bringing the total vocabulary size to 258. This processing approach effectively structures the data for learning, capturing intricate patterns in packet transmissions.

\subsection{Performance of Byte Transformer}
\noindent

To optimize data compression performance while maintaining computational efficiency, ByteTrans incorporates model adaptation by training four versions of predictive model, with parameter sizes of 0.5M, 5M, 55M, and 103M, respectively. Each model is specifically designed to operate within the computational constraints of different devices, ranging from powerful servers to resource-limited edge devices. 
Detailed specifications of each model's configuration are provided in Table \ref{para}.
The model training is conducted on a Thinkstation equipped with an NVIDIA RTX A6000 GPU, using byte packets with a sequence length of 256. Batch sizes are adjusted according to each model's complexity to optimize memory usage and minimize training times. 

\begin{table}[t]
\centering
\caption{Detailed model parameters.}
\begin{tabular}{c|cccc}
\toprule
\textbf{Model Size} & $\mathbf{N_h}$ & $\mathbf{N_{Layer}}$ & \textbf{Dimension} & \textbf{Batch Size} \\
\midrule
0.5M & 8 & 3 & 32 & 256 \\
5M & 8 & 6 & 128 & 256 \\
55M & 8 & 12 & 512 & 128 \\
103M & 12 & 12 & 768 & 64 \\
\bottomrule
\end{tabular}
\label{para}
\end{table}

\begin{figure}[t]  
    \centering
    \begin{subfigure}{0.49\linewidth}  
        \centering
        \includegraphics[width=\linewidth]{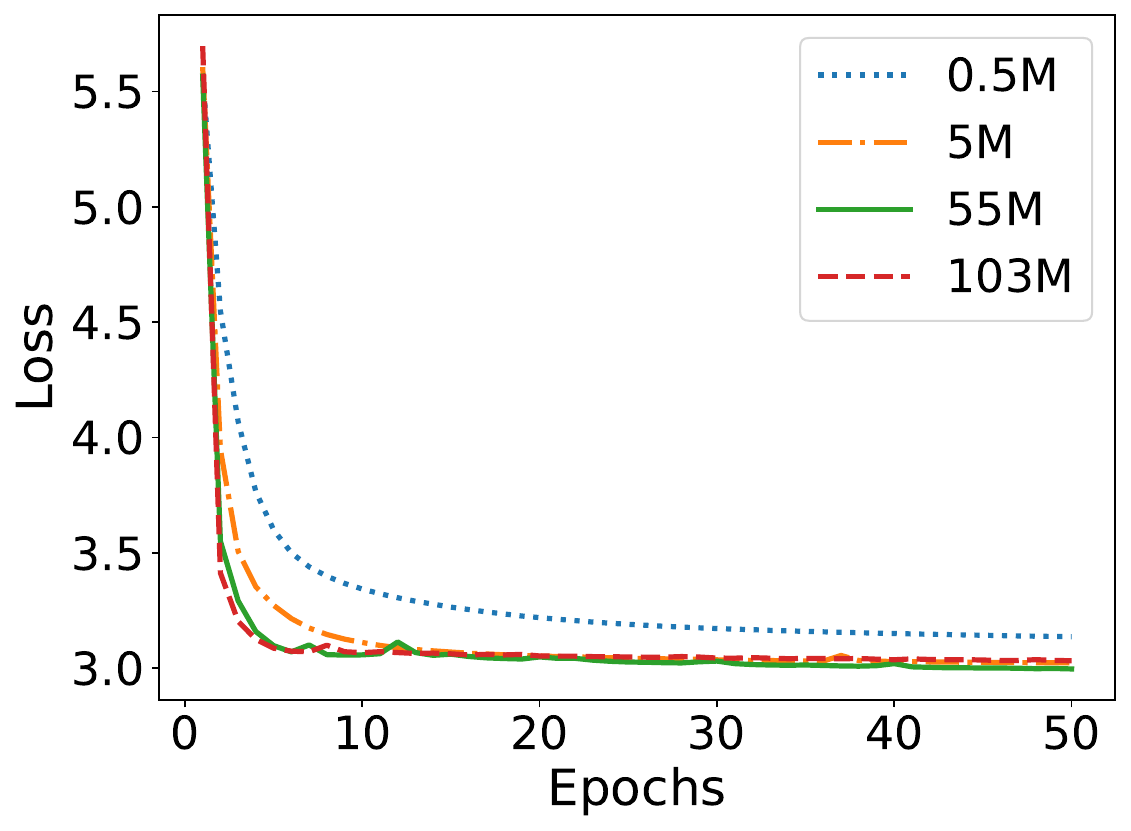}
        \caption{Training loss across different model sizes.} 
	\label{loss}
    \end{subfigure}
    \hfill
    \begin{subfigure}{0.46\linewidth}
        \centering
        \includegraphics[width=\linewidth]{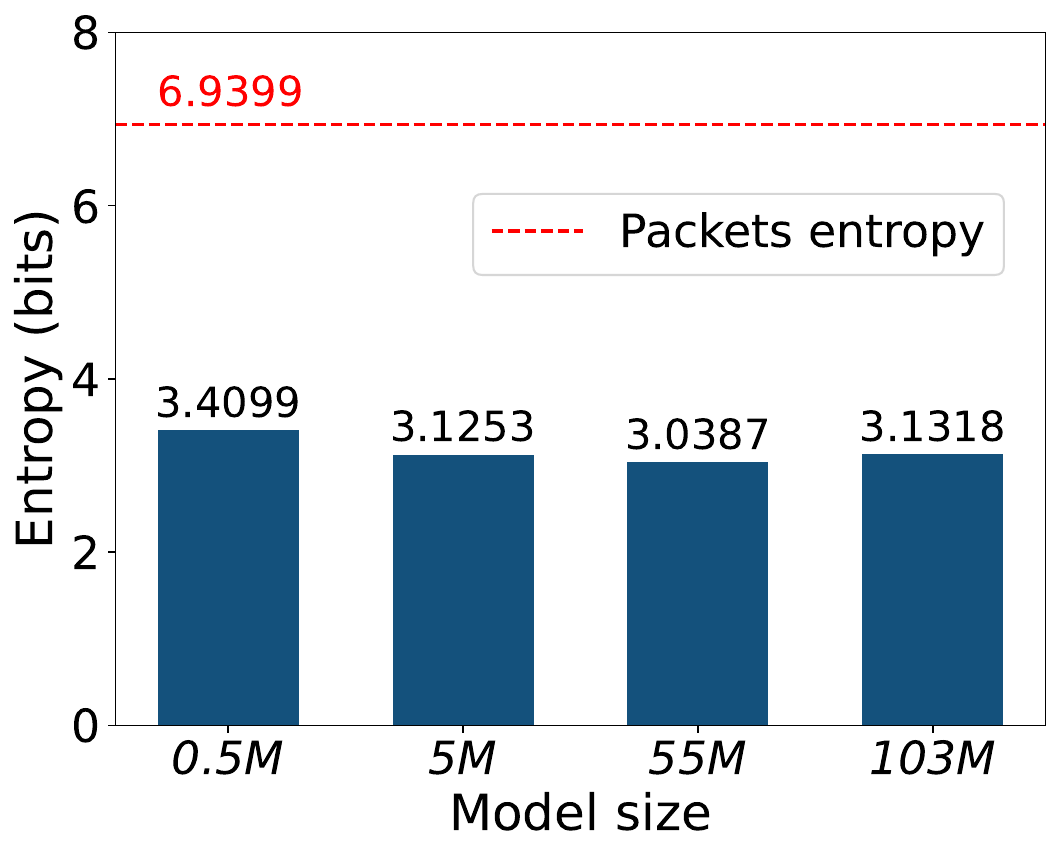}
        \caption{Comparison of entropy values for different models.} 
	\label{Entropy}
    \end{subfigure}
    \caption{Performance of the predictive models.} 
	\label{trans}
    \vspace{-5mm}
\end{figure}

Figure \ref{loss} presents the training loss across various epochs for predictive models of differing sizes utilized in ByteTrans. To delineate the training dynamics for the four models, it depicts the training loss trajectories with a decrement across epochs, signifying effective learning and optimization over time. 
Notably, the larger models exhibit more substantial reduction in loss, achieving rapid convergence when compared to 0.5M and 5M models. This trend suggests that larger models possess an enhanced capacity to assimilate and learn from the data, reaching lower loss values in fewer epochs. In contrast, the smaller models exhibit a slower decrease in loss, reflecting limitations in capturing the data complexity.

Next, we analyze the distribution of the top-10 ranks produced by the predictive models of varying sizes. The results, depicted in Table \ref{table_example}, highlight the \textit{concentration} of ranks, demonstrating the rank transformation capability achieved by each model. In Table \ref{table_example}, '\textbf{R}' represents the rank values assigned to the bytes, and '\textbf{\%}' indicates the percentage of bytes assigned to each rank. Notably, for rank '0', which represents the highest predictability, the 0.5M model assigned this rank to 66.33\% of the data sequences. This percentage increased with the model size, with the 5M and 55M models assigning rank '0' to 70.24\% and 71.05\% of the sequences, respectively.
For the larger 103M model, a slight decrease in the frequency of rank '0' to 69.98\% compared to the 55M model suggests potential overfitting, where the model's larger capacity might be overly tuned to the training data, thereby slightly impairing its generalization ability.
This pattern highlights the balance needed when scaling models, 
which necessitates careful calibration of model capacity to match data diversity and complexity. The steady decrease in percentage from the highest to lower ranks across all models underlines the ability of ByteTrans to effectively predict a large portion of the data with high certainty, leading to efficient data compression by minimizing variability within the predicted ranks.
\begin{table}[t]
\centering
\caption{Top-10 ranks of different predictive models.}
\begin{tabular}{cccccccc}
\toprule
\multicolumn{2}{c}{\textbf{0.5M}} & \multicolumn{2}{c}{\textbf{5M}} & \multicolumn{2}{c}{\textbf{55M}} & \multicolumn{2}{c}{\textbf{103M}}\\
\cmidrule(r){1-2} \cmidrule(lr){3-4} \cmidrule(l){5-6} \cmidrule(l){7-8}
\textbf{R} & \bm{\%} & \textbf{R} & \% & \textbf{R} & \% & \textbf{R} & \%\\
\midrule
\textbf{0} & \textbf{66.33\% }& \textbf{0} & \textbf{70.24\%} & \textbf{0} & \textbf{71.05\%} & \textbf{0} & \textbf{69.98\%}\\
1 & 2.78\% & 1 & 1.87\% & 1 & 1.99\% & 1 & 2.24\%\\
2 & 1.52\% & 2 & 1.06\% & 2 & 1.08\% & 2 & 1.12\%\\
3 & 1.01\% & 3 & 0.88\% & 3 & 0.92\% & 3 & 0.89\%\\
4 & 0.88\% & 4 & 0.69\% & 6 & 0.69\% & 4 & 0.71\%\\
6 & 0.85\% & 6 & 0.62\% & 4 & 0.68\% & 5 & 0.68\%\\
5 & 0.82\% & 5 & 0.62\% & 5 & 0.64\% & 6 & 0.60\%\\
8 & 0.58\% & 7 & 0.58\% & 7 & 0.61\% & 7 & 0.50\%\\
7 & 0.58\% & 9 & 0.49\% & 8 & 0.52\% & 10 & 0.50\%\\
9 & 0.48\% & 8 & 0.48\% & 9 & 0.43\% & 9 & 0.43\%\\
\bottomrule
\end{tabular}
\label{table_example}
\vspace{-2mm}
\end{table}

To further quantify the uncertainty inherent in the data sequence, Figure \ref{Entropy} illustrates the entropy values of the transformed rank-based sequences generated by the predictive models, compared to the entropy of the original packet data. The original data sequences have an entropy of 6.9399 bits, serving as a benchmark. It is observed that as model size increases, the entropy decreases with the 0.5M, 5M, and 55M models, demonstrating enhanced data pattern capture and predictability. However, the 103M model shows a slight increase in entropy, suggesting potential overfitting and reduced efficiency in assigning lower rank values, consistent with the rank distribution results in Table~\ref{table_example}. This trend underscores the importance of optimal model sizes to balance compression efficiency and overfitting risks.

\subsection{Performance of Data Compression}
\noindent

Integrated with the lossless compression method in our ByteTrans, Table \ref{tab:example} presents the experimental results of data compression capability across models of different sizes. This result captures the quantitative performance of three predictive models, each calibrated to distinct parameter scales and evaluated based on their ability to compress network packets under controlled conditions.
In this evaluation, we divide the test dataset into 10 distinct groups to systematically assess the compression performance. Specifically, \textit{Group No.} represents different batches of data packets being processed, while \textit{$N_P$} and \textit{$N_B$} indicate the number of packets and the total number of corresponding bytes in each group, respectively. The columns \textit{$S_{0.5M}$}, \textit{$S_{5M}$}, \textit{$S_{55M}$}, and \textit{$S_{103M}$} represent the compressed sizes achieved by each predictive model. 
Notably, we observe that larger models generally achieving better compression ratio across most groups. However, the 103M model exhibits a slight increase in compressed size compared to the 55M model, suggesting diminishing returns as model size increases. This also highlights the 5M model as a balanced option between compression performance and computational demand.

\begin{figure}[t] 
    \centering
    \begin{subfigure}{0.48\linewidth}  
        \centering
        \includegraphics[width=\linewidth]{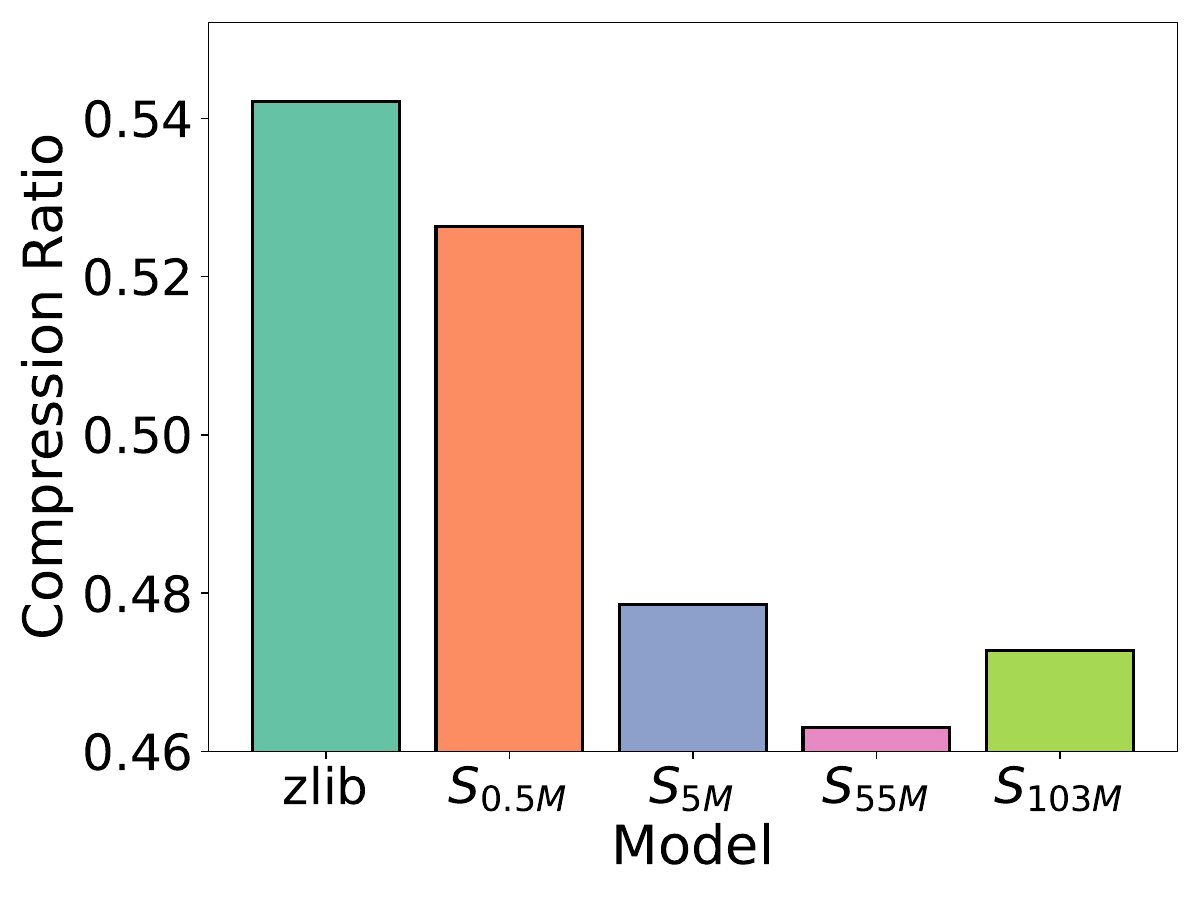}
        \caption{compression ratio for different models.}
        \label{rate1}
    \end{subfigure}
    \hfill
    \begin{subfigure}{0.47\linewidth}
        \centering
        \includegraphics[width=\linewidth]{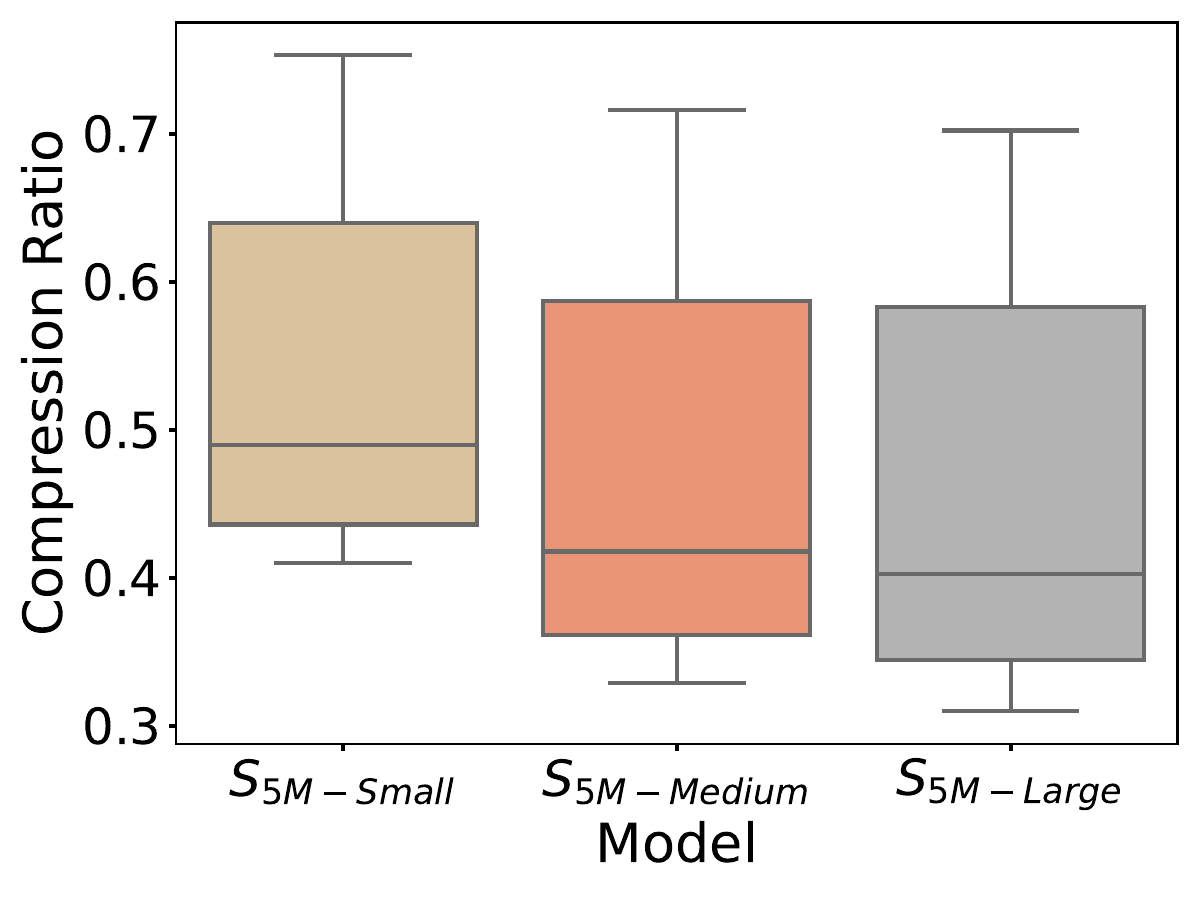}
        \caption{compression ratio for different training data sizes.}
        \label{rate2}
    \end{subfigure}
    \caption{compression ratio under various conditions.}
    \label{rate}
\end{figure}

\begin{table}[t]
\centering
\caption{Compression performance of different model sizes.}
\begin{tabular}{c|c|cccc}
\toprule
\textbf{Group No.} & \bm{$N_B$} & \bm{$S_{0.5M}$} & \bm{$S_{5M}$} & \bm{$S_{55M}$} & \bm{$S_{103M}$} \\
\midrule
1 & 24,422 & 9,086 & 8,025 & \textbf{7,426} & 7,838\\
2 & 22,905 & 11,893 & 10,327 & \textbf{10,113} & 10,280\\
3 & 28,228 & 16,266 & 14,419 & \textbf{14,071} & 14,355\\
4 & 24,708 & 10,160 & 9,196 & \textbf{8,761} & 8,908\\
5 & 25,343 & 10,109 & 9,062 & \textbf{8,630} & 8,860\\
6 & 24,807 & 10,249 & 8,683 & \textbf{8,275} & 8,647\\
7 & 27,869 & 18,116 & 17,213 & \textbf{16,766} & 17,012\\
8 & 30,176 & 22,342 & 21,606 & \textbf{21,211} & 21,392\\
9 & 26,219 & 17,661 & 16,066 & \textbf{15,934} & 16,232\\
10 & 24,350 & 10,440 & 9,366 & \textbf{8,766} & 8,932\\
\midrule
\textbf{Total} & 259,027 & 136,322 & 123,963 & \textbf{119,953} & 122,456\\
\bottomrule
\end{tabular}
\label{tab:example}
    \vspace{-5mm}
\end{table}

\begin{table*}[ht]
\vspace{2mm}
\centering
\renewcommand{\arraystretch}{1.3} 
\caption{Hardware platform specifications.}
\begin{tabular}{c|ccccc}
\toprule
\textbf{Device Type} & CPU & GPU Type & Graphic Memory & Power & RAM \\
\midrule
Thinkstation P620 & AMD Ryzen 3945WX (12-core) & NVIDIA RTX A5000 & 24GB & 230W$_{(GPU)}$ & 256GB \\
Jetson Orin Nano & ARM Cortex-A57 (4-core) & NVIDIA Ampere architecture GPU & 8GB & 15W & 8GB \\
Raspberry Pi 4B & ARM Cortex-A72 (4-core) & \ding{55} & \ding{55} & 15W & 4GB \\
\bottomrule
\end{tabular}
\label{testbed}
\vspace{-3mm}
\end{table*}

Figure \ref{rate1} shows the data compression ratio for different ByteTrans model variants and the baseline model zlib \cite{zlib}. 
As expected, it is observed that the mean compression ratios decrease as the model size increases, indicating that larger models are more effective at compression due to their enhanced ability to capture complex data patterns. However, even with smaller models, such as the 5M or 0.5M versions, the compression ratio still ranges from 47.8\% to 52.6\%, indicating that the data size after lossless compression is less than half of the original packet size. This demonstrates that ByteTrans performs effectively even with the smallest model.
Similarly, the $S_{103M}$ model shows a slight increase in compression ratio compared to the $S_{55M}$ model, suggesting diminishing returns at this larger scale. Notably, the performance of the $S_{5M}$ model is comparable to that of the $S_{55M}$ model, which demonstrates a favorable balance between model efficiency and compression effectiveness. This underscores a practical trade-off between computational efficiency and accuracy, supporting the broader deployment of the smaller $S_{5M}$ model for its cost-effective performance.
Notably, all our models outperform the baseline compression library, zlib, with compression ratio improvements ranging from 11.8\% (5M) to 14.6\% (55M), demonstrating the superior ability of ByteTrans to compress data more efficiently.

Furthermore, we investigate the impact of varying training dataset sizes on compression performance, with a particular focus on the $S_{5M}$ model. As shown in Figure \ref{rate2}, the compression ratio decreases as the size of the training dataset increases, indicating that a larger dataset enables the model to make more accurate predictions. By contrast, the results show a significant improvement in compression ratio when increasing the dataset size from small to medium. However, beyond this point, further increases in the dataset size yield diminishing returns in performance improvement.
While larger datasets improve accuracy, they come at the cost of increased training time and resource demands, making it critical to balance training efficiency with optimal compression performance.

\subsection{Implementation Efficiency}
\noindent

\begin{figure}[t]
    \centering
    
    \begin{subfigure}[b]{0.15\textwidth}
        \centering
        \includegraphics[width=\textwidth]{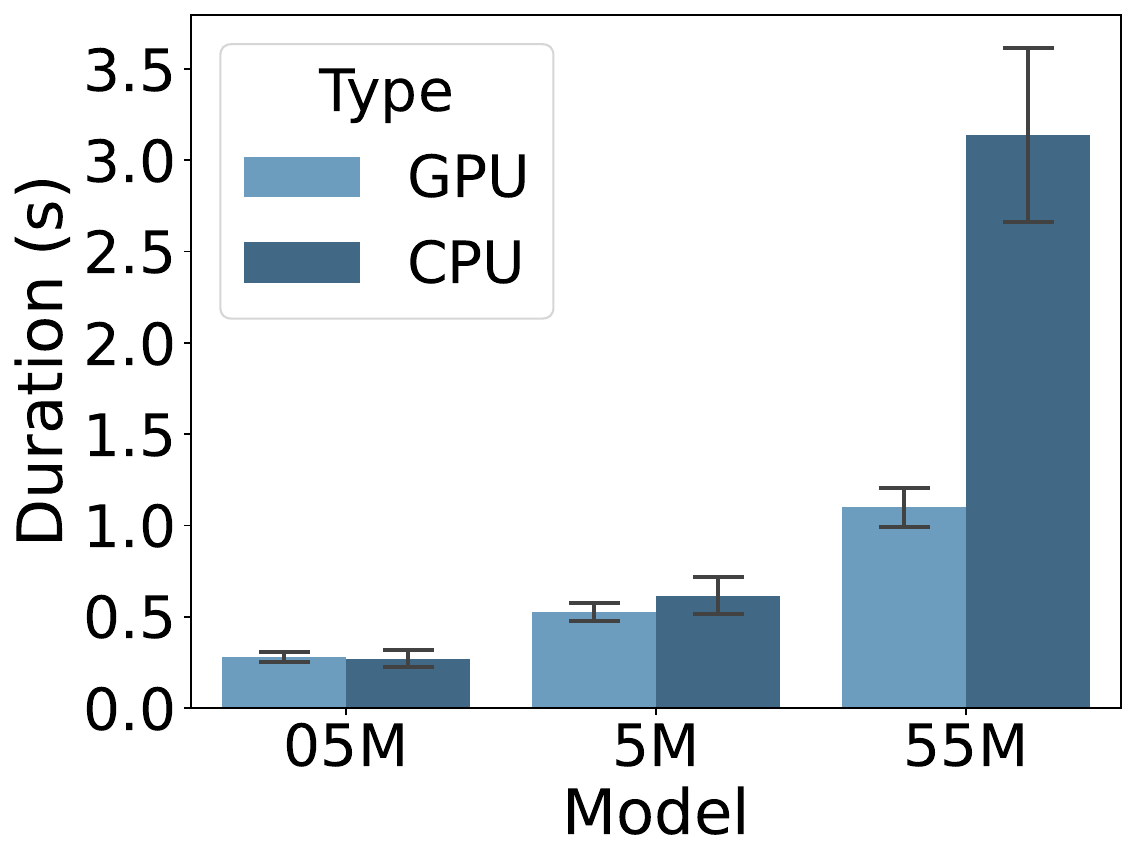}
        \caption{A5000 Server}
        \label{servertime}
    \end{subfigure}
    \hfill
    \begin{subfigure}[b]{0.15\textwidth}
        \centering
        \includegraphics[width=\textwidth]{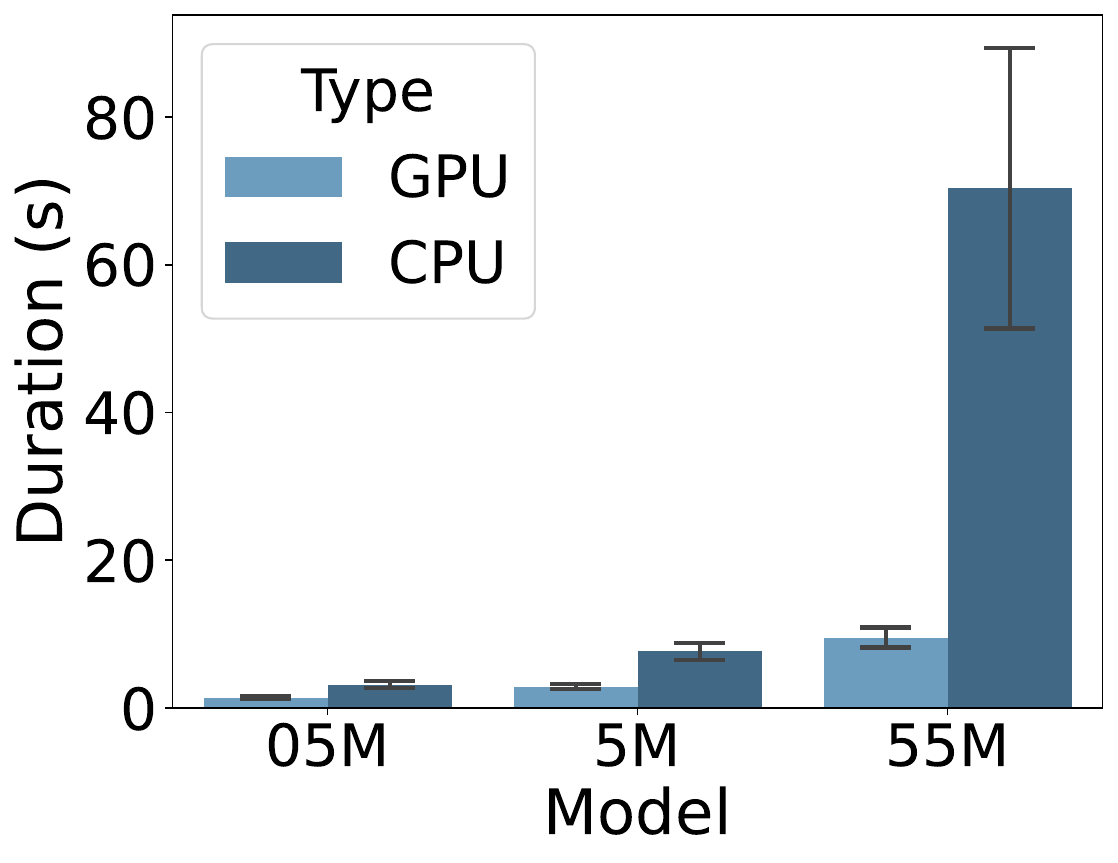}
        \caption{Jetson Nano}
        \label{jetsontime}
    \end{subfigure}
    \hfill
    \begin{subfigure}[b]{0.15\textwidth}
        \centering
        \includegraphics[width=\textwidth]{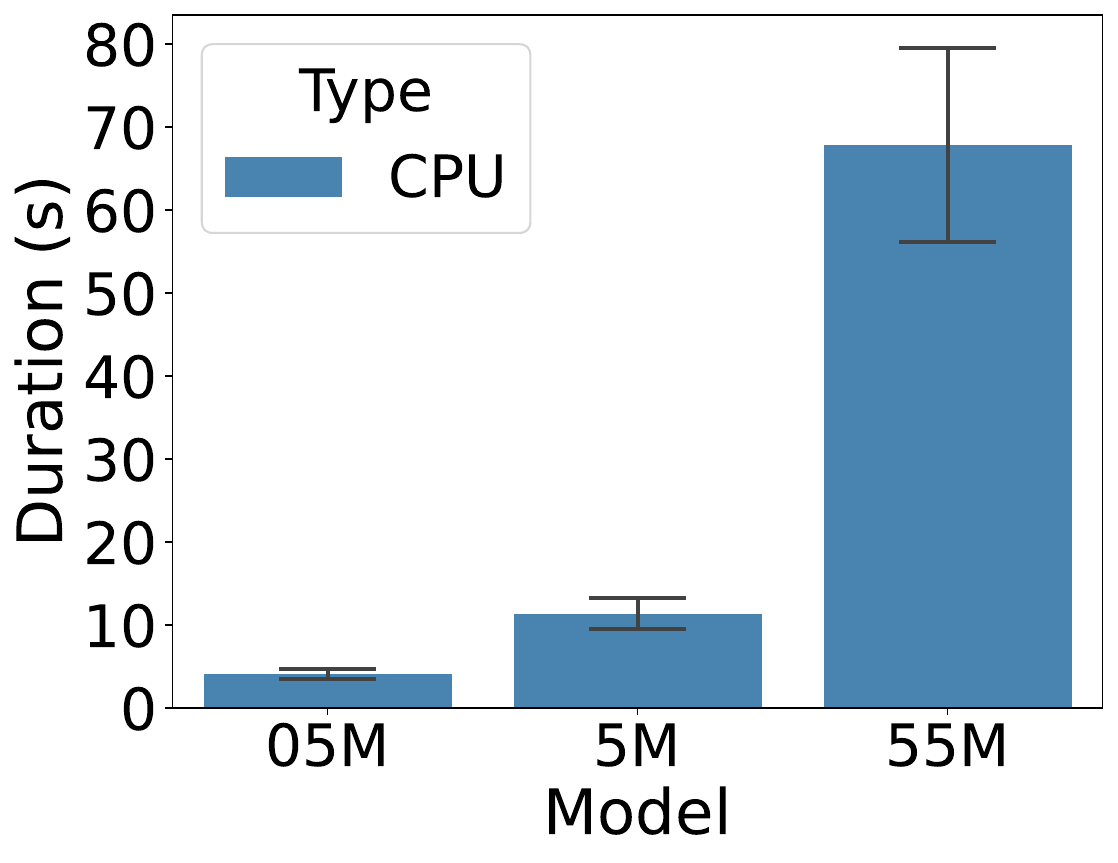}
        \caption{Raspberry Pi}
        \label{pitime}
    \end{subfigure}
    \caption{Compression time on different devices.}
    \label{fig:three_images}
    \vspace{-6mm}
\end{figure}

To assess the practical applicability of ByteTrans on a variety of IoT hardware platforms, we conduct a series of real-world deployments and evaluations in multiple dimensions. 
Specifically, we test the designed models on Raspberry Pi devices, Nvidia Jetson modules, and high-performance (HP) servers, which are commonly used IoT applications, to cover a broad spectrum of computational capabilities. The detailed hardware  specifications can be found in Table \ref{testbed}.
We directly benchmark ByteTrans on the PyTorch implementation.
For the Raspberry Pi and the HP server, we utilize the \texttt{psutil} library to obtain resource utilization information for the data compression process. 
We also employ the \texttt{nvidia-smi} API and \texttt{tegrastats} utility to monitor resource usage during network packet compression with different devices. 
These benchmark operations run on a separate process, minimizing interference with the compression process in real-world environments.

\begin{figure}[t]
    \centering
    \begin{subfigure}[b]{0.23\textwidth}
        \centering
        \includegraphics[width=\textwidth]{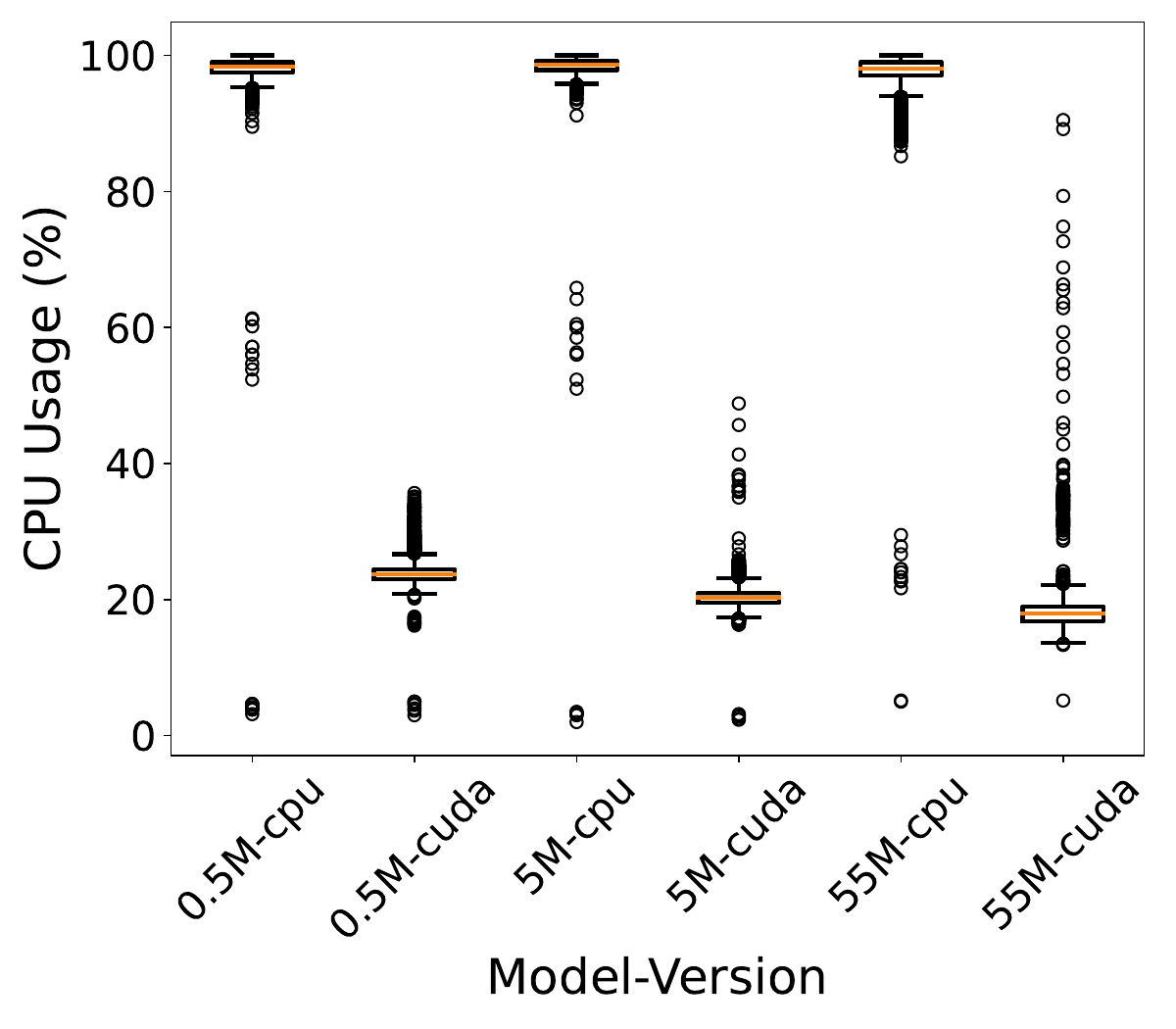}
        \caption{CPU}
        \label{cpu}
    \end{subfigure}
    \hfill
    \begin{subfigure}[b]{0.23\textwidth}
        \centering
        \includegraphics[width=\textwidth]{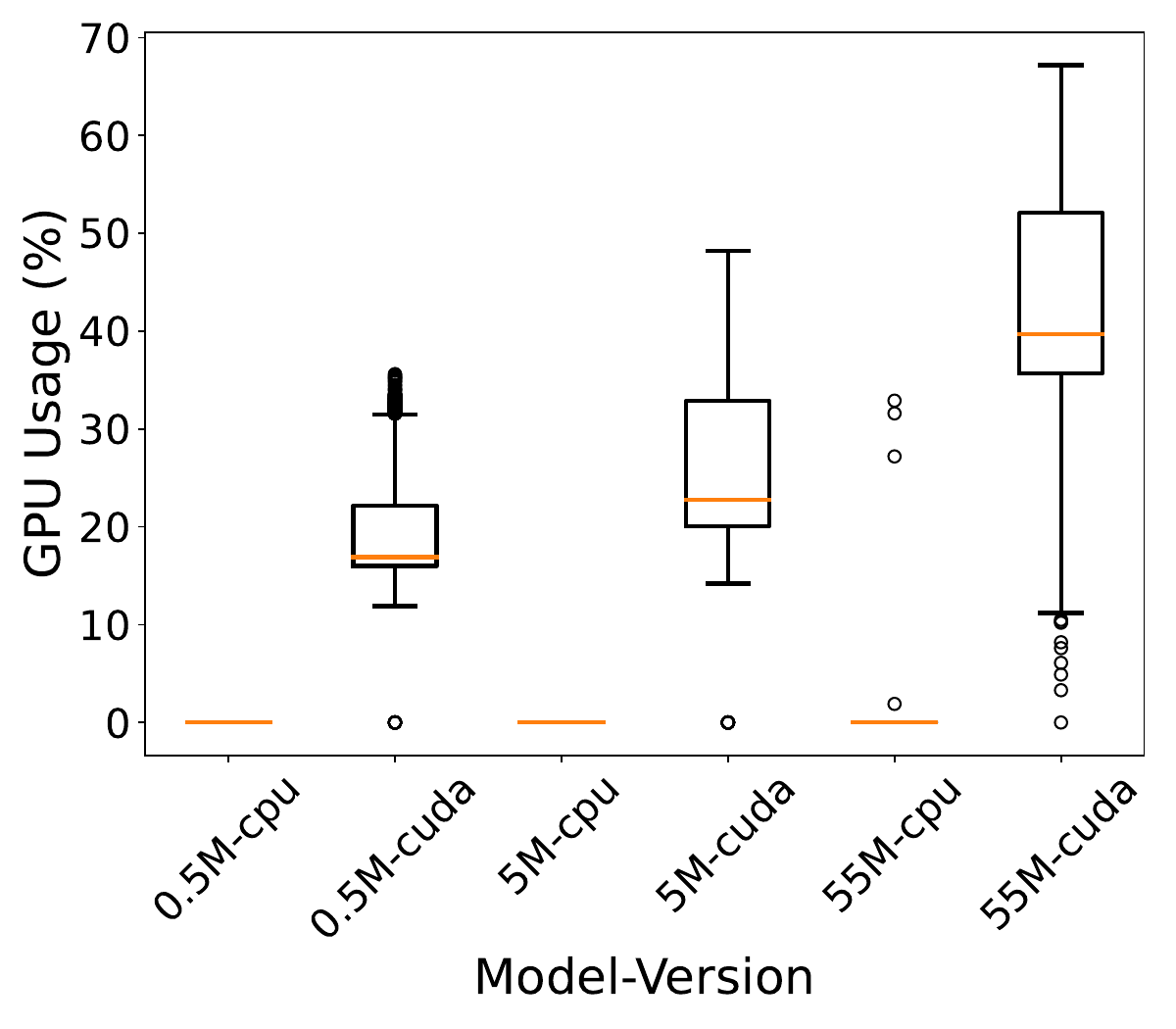}
        \caption{GPU}
        \label{gpu}
    \end{subfigure}

    \begin{subfigure}[b]{0.23\textwidth}
        \centering
        \includegraphics[width=\textwidth]{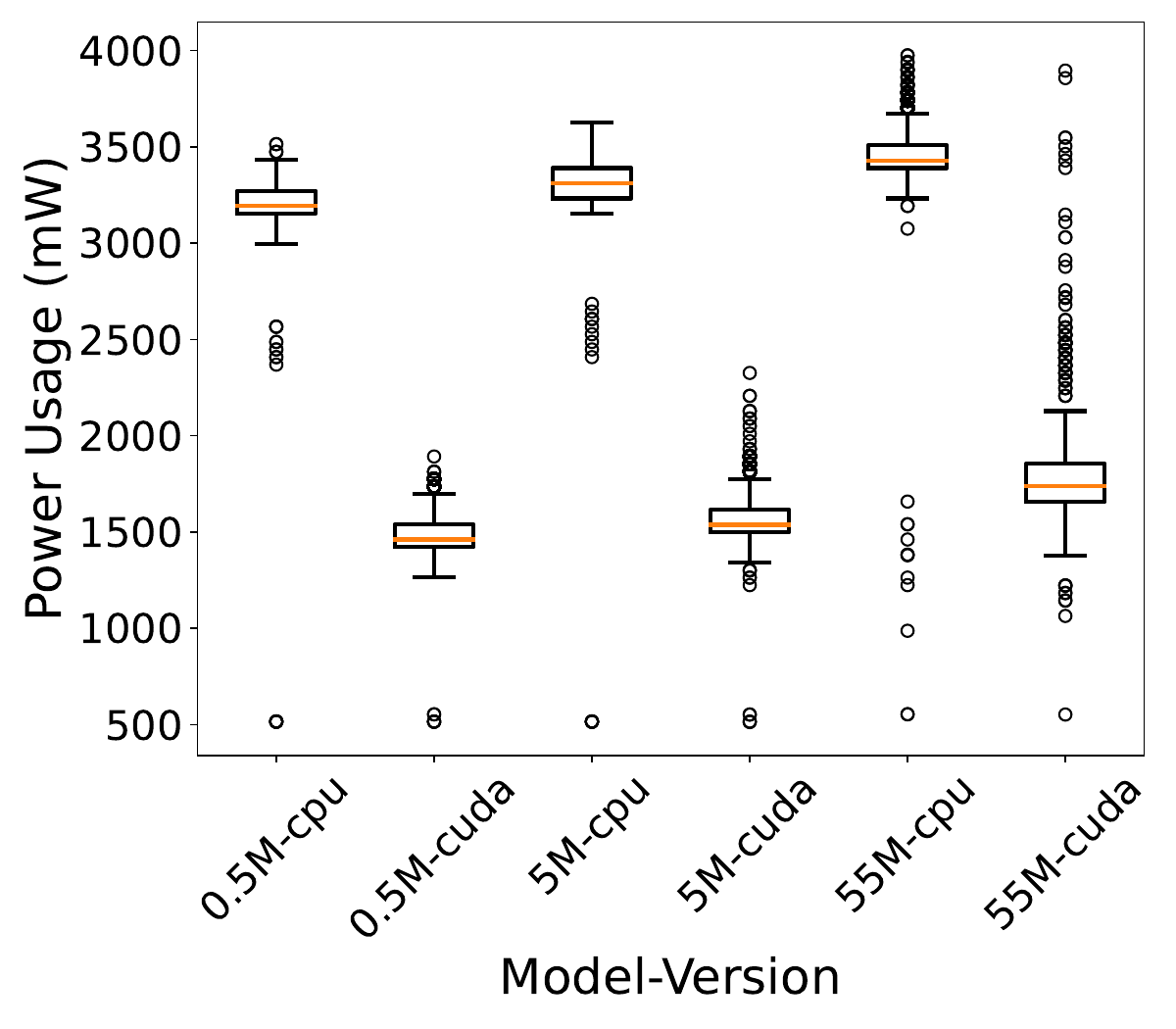}
        \caption{Overall Power}
        \label{power}
    \end{subfigure}
    \hfill
    \begin{subfigure}[b]{0.23\textwidth}
        \centering
        \includegraphics[width=\textwidth]{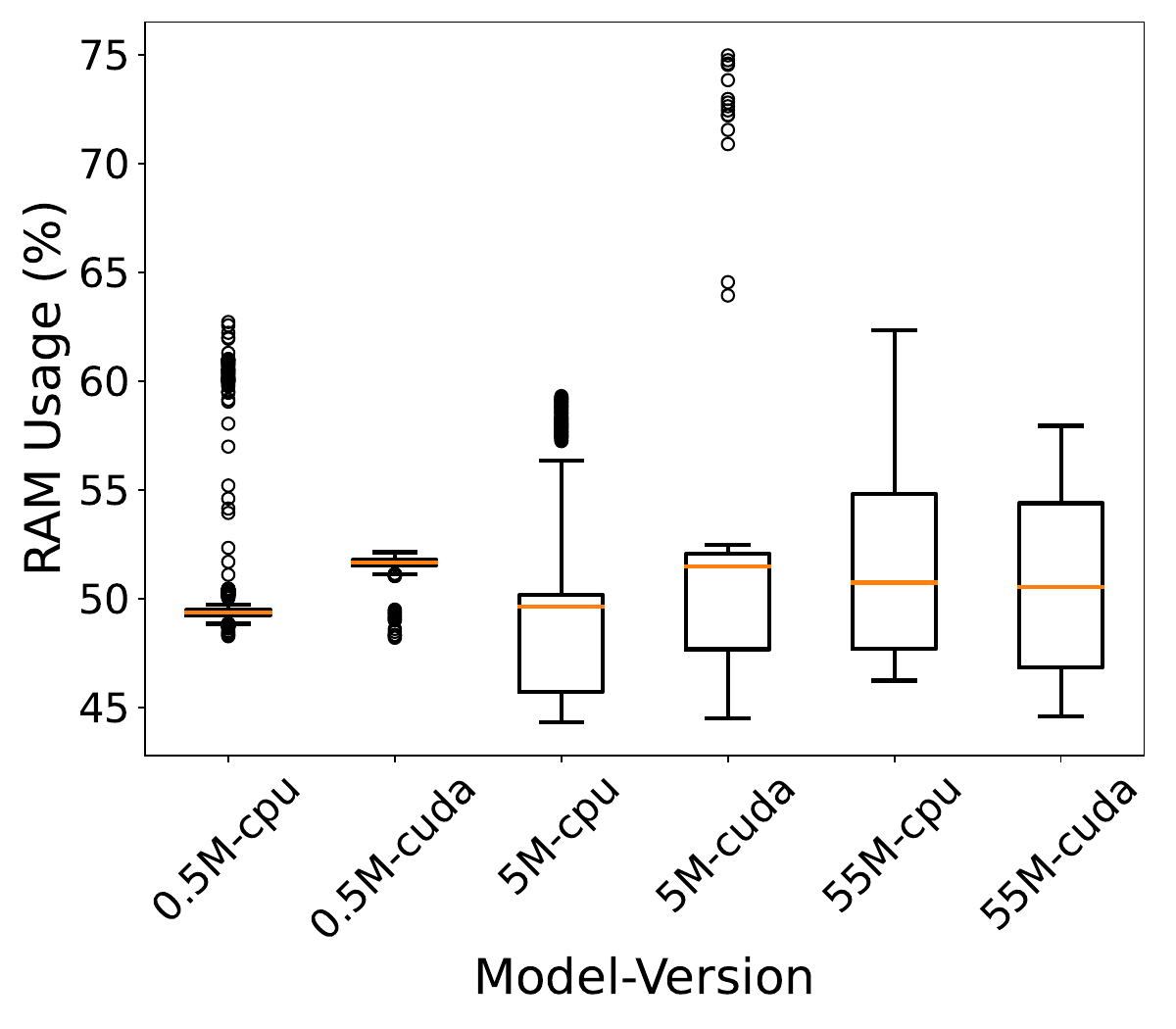}
        \caption{RAM}
        \label{ram}
    \end{subfigure}

    \caption{Resource usage of the edge device.} 
    \label{jetson}
    \vspace{-6mm}
\end{figure}

\subsubsection{Compression Time}
\noindent
The evaluation of compression times across different devices reveals key insights into the balance between model size and available computational resources. Notably, GPU-accelerated models demonstrate a significant performance advantage over their CPU counterparts across all platforms, particularly with larger models like the $S_{55M}$ performing notably slower on CPU. This trend of increasing compression time with the larger models is consistent across all devices tested. On the server side, all models achieve faster data compression, benefiting from the server's robust processing capabilities. On the edge devices, where computational resources are more constrained, the performance disparity between GPU and CPU processing becomes more pronounced. This emphasizes the effectiveness of using smaller, resource-efficient models like the $S_{5M}$, which achieves a compression ratio nearly comparable to the $S_{55M}$, offering a favorable tradeoff between performance and resource usage. 
In IoT scenarios, communication entities typically collect and transmit data to a central node or gateway at predefined intervals \cite{zhang2022aoi}. This periodic communication, which can range from seconds to hours depending on the application and sensor design, provides ample time to compress packets. As a result, the compression time of our models is well-suited for such environments where periodic communication is standard.

\begin{figure}[t]
    \centering
    \begin{subfigure}[b]{0.23\textwidth}
        \centering
        \includegraphics[width=\textwidth]{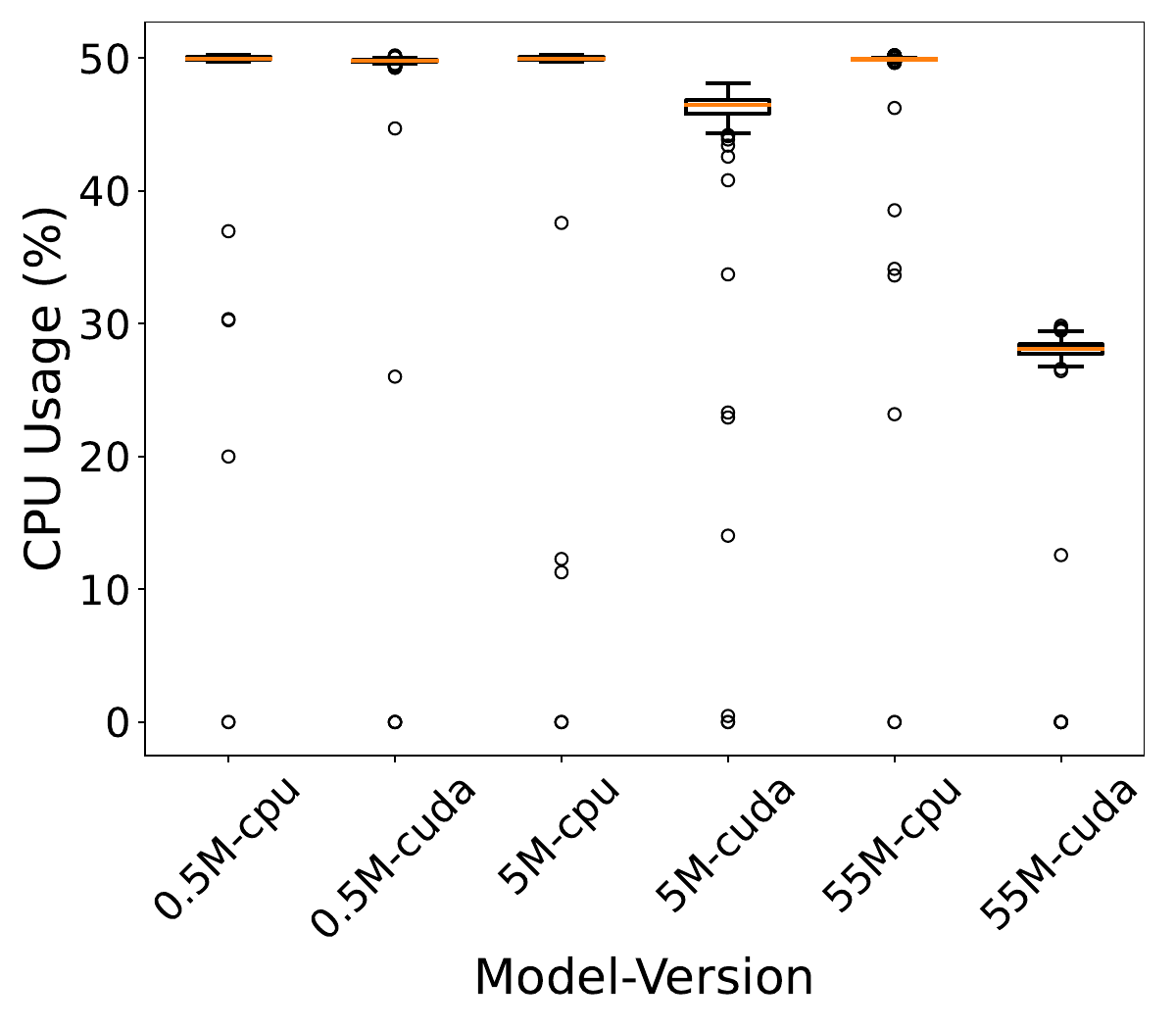}
        \caption{CPU}
        \label{servercpu}
    \end{subfigure}
    \hfill
    \begin{subfigure}[b]{0.23\textwidth}
        \centering
        \includegraphics[width=\textwidth]{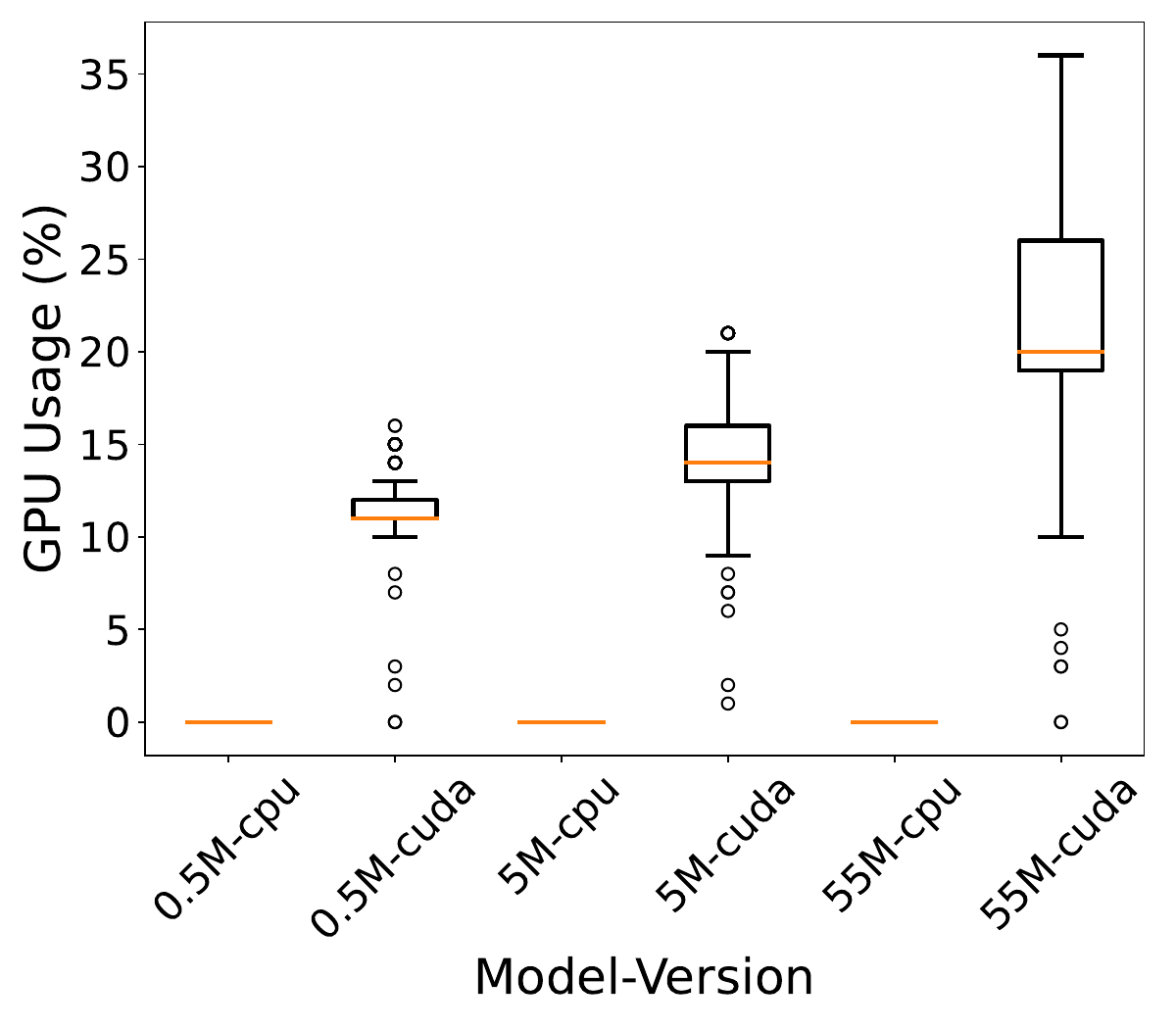}
        \caption{GPU}
        \label{servergpu}
    \end{subfigure}

    \begin{subfigure}[b]{0.23\textwidth}
        \centering
        \includegraphics[width=\textwidth]{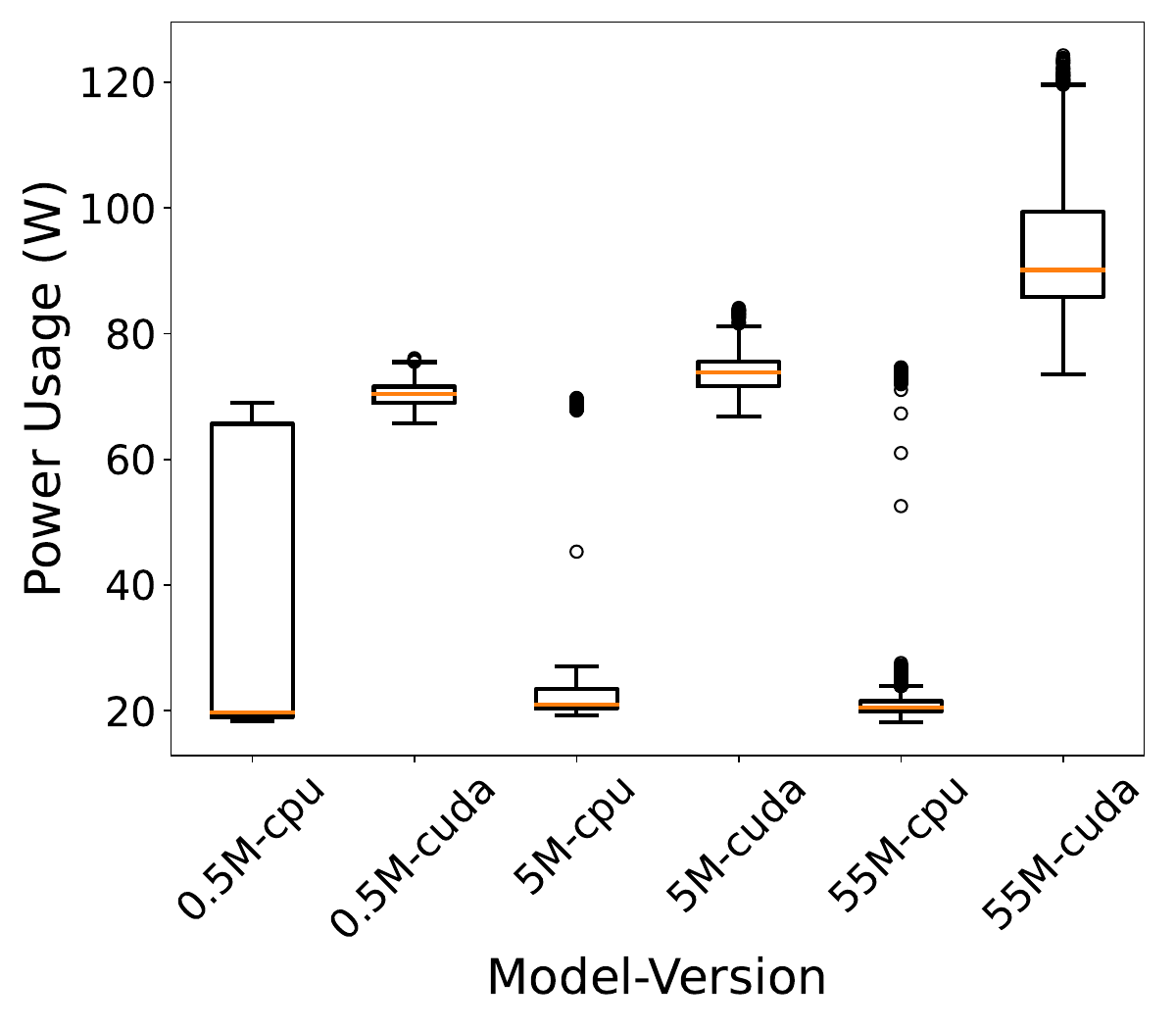}
        \caption{GPU Power}
        \label{serverpower}
    \end{subfigure}
    \hfill
    \begin{subfigure}[b]{0.23\textwidth}
        \centering
        \includegraphics[width=\textwidth]{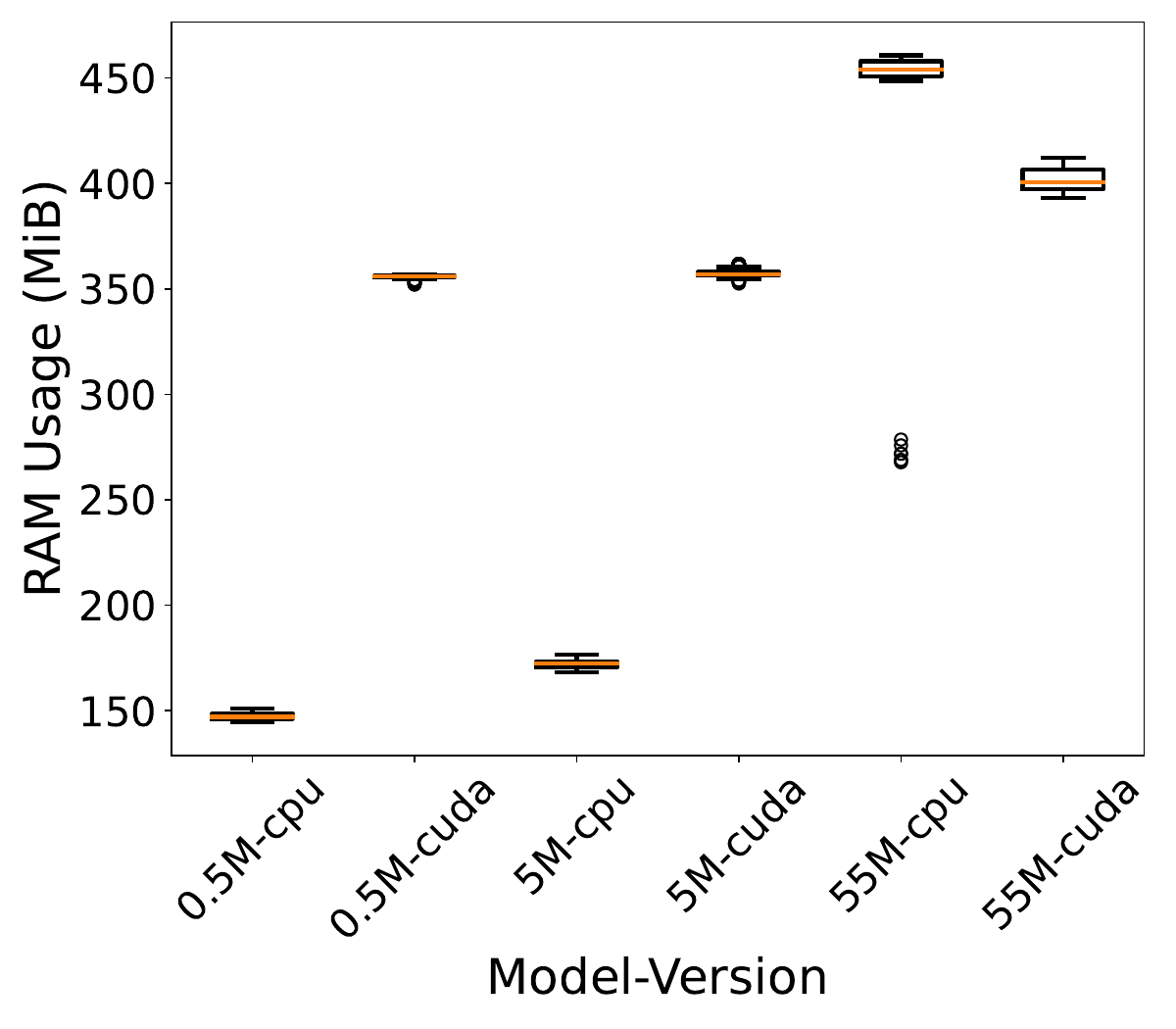}
        \caption{RAM}
        \label{serverram}
    \end{subfigure}

    \caption{Resource usage of HP server.}
    \label{server}
    \vspace{-7mm}
\end{figure}

\subsubsection{Resource Usage}
\noindent
In the context of resource utilization, we conduct the detailed evaluation of different model versions on various platforms to investigate their actual resource demands. Figure \ref{jetson} illustrates the resource consumption across several metrics for models deployed on the Jetson device.
The resource utilization across different models on the Jetson device reveals distinct trends in CPU and GPU usage, power consumption, and RAM utilization. The CPU usage, as depicted in Figure \ref{cpu}, shows a marked difference between CPU-only and GPU-accelerated versions of the models. 
The CPU-only versions of the models exhibit high CPU usage, approaching 100\%, while GPU-accelerated versions significantly reduce CPU load,  demonstrating notably lower CPU usage across all model sizes.
It is worth noting that the larger-size model will incur more occasional high CPU usage.
This might be due to the intensive GPU operation orchestration or data loading and can lead to future investigations.
In contrast, the GPU usage shown in Figure \ref{gpu} is minimal for CPU versions of the models but escalates in GPU-accelerated versions. 
Power consumption, detailed in Figure \ref{power}, shows that the larger-size models require more power, particularly with CPU versions generally consuming more than their GPU counterparts. However, adopting GPU for inference can significantly reduce the power consumption, which indicates that employing an edge GPU not only accelerate the data compression efficiency but also improve the power efficiency.
Lastly, Figure~\ref{jetson}\subref{ram} illustrates the Random Access Memory (RAM) usage for various model versions. 
Notably, increasing the model size does not significantly increase RAM overhead. 
These observations highlight the broad compatibility of our ByteTrans across various model sizes on edge devices. Nevertheless, we suggest conducting profiling before deployment to select the most appropriate model size based on specific communication requirements and available device resources.

The benchmark results on the Thinkstation server are presented in Figure~\ref{server}.
We record the CPU usage, GPU usage, GPU runtime power consumption, and RAM usage of the compression process.
In Figure~\ref{servertime}, we observe that when inferring with CPU, the consumption of models with different sizes is similar. 
This demonstrates the negligible computational overhead when applying larger models on the devices with abundant computational resources.
Meanwhile, we notice that when performing inferences on GPU, the CPU consumption decreases as the model size increases. 
This is because, when inferring on GPU, the major task undertaken by CPU cores is data loading.
As shown in Figure~\ref{servertime}, a larger model takes more time to make predictions for data compression, thus lowering the frequency of data loading and decreasing CPU consumption.
From Figures \ref{servergpu} and \ref{serverpower}, it is observed that when inferring with GPU, the models show significantly higher GPU usage and power consumption than when inferring with CPU, with these values escalating as model size increases. This underscores the increased computational demands imposed by larger models on GPU resources. Similarly, RAM usage also increases with model size as shown in Figure \ref{serverram}, demonstrating a proportional increase in resource allocation. 
It is worth noting that the RAM utilization when inferring with GPU are higher than with CPU for small models. 
This is because our small models do not consume much memory, and inferring with GPU requires more frequent data exchange between RAM and graphic memory.
Considering the similar inference time on rank sequences when running on two devices as shown in Fig.~\ref{fig:three_images}\subref{servertime}, such a phenomenon indicates deploying small models on GPU does not have obvious advantages while may introduce additional overhead.

\begin{figure}[t]  
    \centering
    \begin{subfigure}{0.48\linewidth}  
        \centering
        \includegraphics[width=\linewidth]{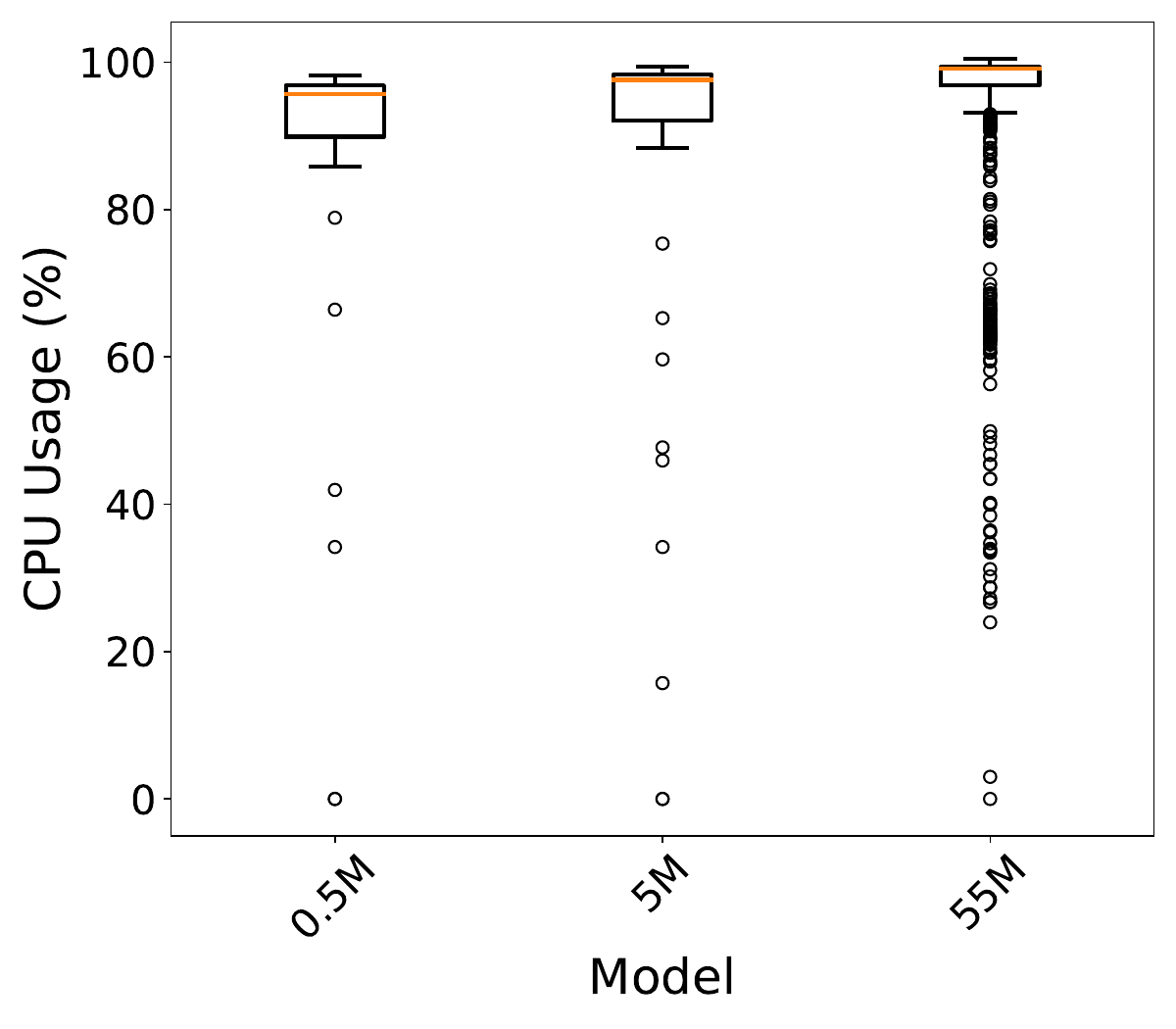}
        \caption{CPU}
        \label{Rasp1}
    \end{subfigure}
    \hfill
    \begin{subfigure}{0.47\linewidth}
        \centering
        \includegraphics[width=\linewidth]{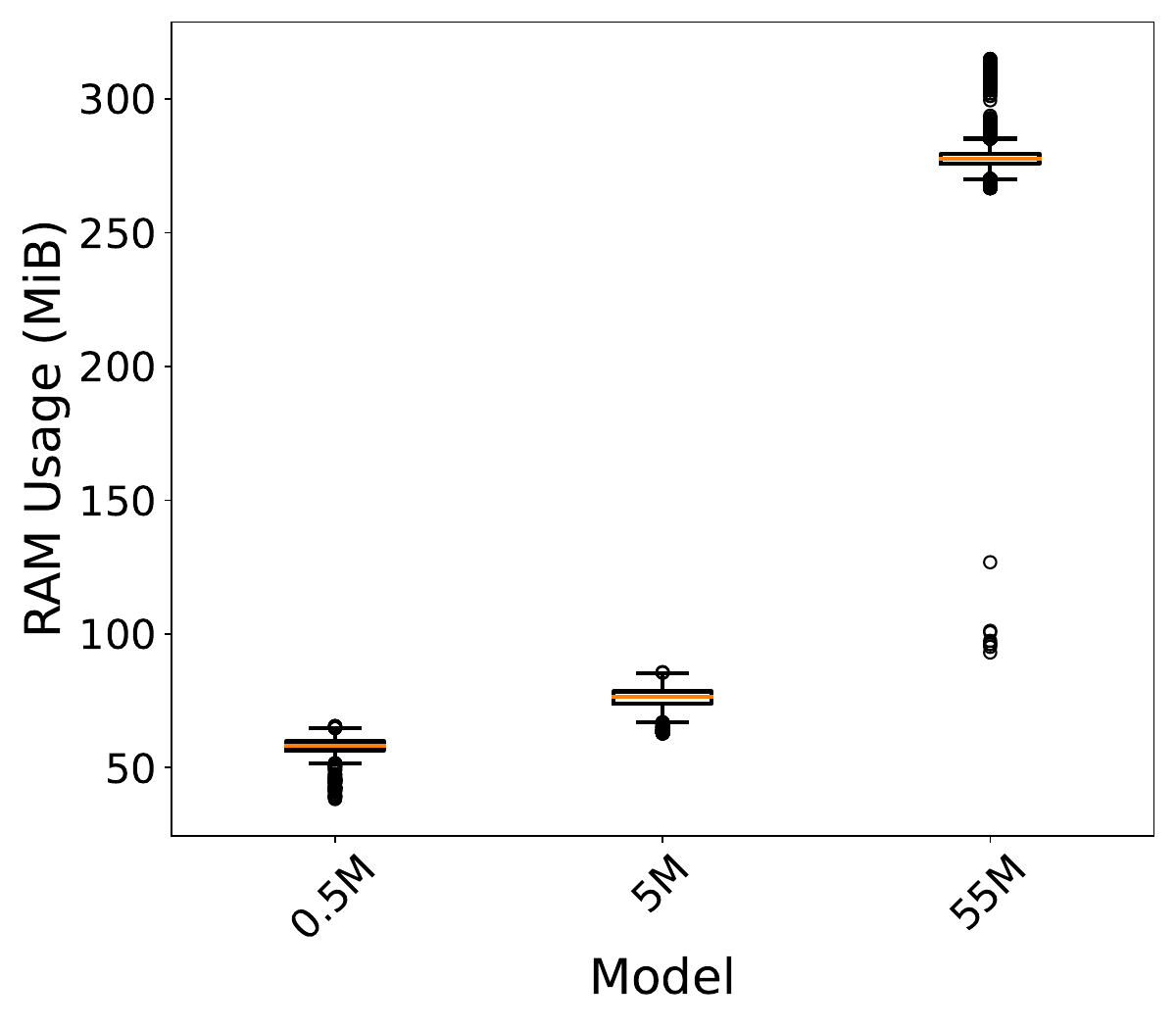}
        \caption{RAM}
        \label{Rasp2}
    \end{subfigure}
    \caption{Resource usage of Raspberry Pi.}
    \label{Rasp}
    \vspace{-6mm}
\end{figure}

Figure \ref{Rasp} illustrates the resource usage on a Raspberry Pi 4B, which lacks GPU capabilities and relies solely on CPU for inference. From Figure \ref{Rasp1}, it is evident that CPU usage remains consistently high across all model sizes, exceeding 90\%, with larger models more extensively utilizing the CPU resource during the inference. Figure \ref{Rasp2} shows RAM usage, highlighting a clear increase as model size grows, indicating more significant memory allocation required for larger models. However, the RAM consumption, even for the largest model, is under 10\% of the total available RAM (4GB for Raspberry Pi). This indicates our ByteTrans models are compatible with most commercial off-the-shelf (COTS) mobile edge devices, including those used in IoT or sensor networks environments, with respect to the memory consumption.

\section{Conclusion}
\noindent
In this study, we introduce a byte-level predictive model for universal packet compression in multi-modal communications, which offers a unified and efficient approach for handling various data types within network systems. This model is particularly adept in environments such as the IoT and sensor networks, where the heterogeneous nature and periodic communication of data necessitate robust and adaptable data compression solutions. Our model not only simplifies the compression process across different data modalities but also ensures substantial data size reduction, which is critical in resource-constrained network environments. 
The evaluation results have demonstrated that ByteTrans can effectively reduce the size of multi-model data by processing them at the byte level, achieving a higher compression ratio compared to the baseline approach. This approach has the potential to set new benchmarks for data compression technology in our interconnected world, where the volume of transmitted data continues to grow exponentially.
{In future work, we will explore model adaptation strategies to dynamically optimize the trade-off between resource usage and compression ratio, adapting to different network conditions and application requirements.}

\section*{Acknowledgment}
{This research was supported by the National Science Foundation through Award CNS--2312138 and CNS--2433966.}

\bibliographystyle{IEEEtran}
\bibliography{reference}

\begin{thebibliography}{10}
\providecommand{\url}[1]{#1}
\csname url@samestyle\endcsname
\providecommand{\newblock}{\relax}
\providecommand{\bibinfo}[2]{#2}
\providecommand{\BIBentrySTDinterwordspacing}{\spaceskip=0pt\relax}
\providecommand{\BIBentryALTinterwordstretchfactor}{4}
\providecommand{\BIBentryALTinterwordspacing}{\spaceskip=\fontdimen2\font plus
\BIBentryALTinterwordstretchfactor\fontdimen3\font minus \fontdimen4\font\relax}
\providecommand{\BIBforeignlanguage}[2]{{%
\expandafter\ifx\csname l@#1\endcsname\relax
\typeout{** WARNING: IEEEtran.bst: No hyphenation pattern has been}%
\typeout{** loaded for the language `#1'. Using the pattern for}%
\typeout{** the default language instead.}%
\else
\language=\csname l@#1\endcsname
\fi
#2}}
\providecommand{\BIBdecl}{\relax}
\BIBdecl

\bibitem{li2015internet}
S.~Li, L.~D. Xu, and S.~Zhao, ``The internet of things: a survey,'' \emph{Information systems frontiers}, vol.~17, pp. 243--259, 2015.

\bibitem{nguyen20216g}
D.~C. Nguyen, M.~Ding, P.~N. Pathirana, A.~Seneviratne, J.~Li, D.~Niyato, O.~Dobre, and H.~V. Poor, ``6g internet of things: A comprehensive survey,'' \emph{IEEE Internet of Things Journal}, vol.~9, no.~1, pp. 359--383, 2021.

\bibitem{sinha2017survey}
R.~S. Sinha, Y.~Wei, and S.-H. Hwang, ``A survey on lpwa technology: Lora and nb-iot,'' \emph{Ict Express}, vol.~3, no.~1, pp. 14--21, 2017.

\bibitem{hsu2017breaking}
S.-H. Hsu, C.-H. Lin, C.-Y. Wang, and W.-T. Chen, ``Breaking bandwidth limitation for mission-critical iot using semisequential multiple relays,'' \emph{IEEE Internet of Things Journal}, vol.~5, no.~5, pp. 3316--3329, 2017.

\bibitem{kumar2018strategy}
S.~Kumar and V.~K. Chaurasiya, ``A strategy for elimination of data redundancy in internet of things (iot) based wireless sensor network (wsn),'' \emph{IEEE Systems Journal}, vol.~13, no.~2, pp. 1650--1657, 2018.

\bibitem{verma2018data}
N.~Verma and D.~Singh, ``Data redundancy implications in wireless sensor networks,'' \emph{Procedia computer science}, vol. 132, pp. 1210--1217, 2018.

\bibitem{shannon1951prediction}
C.~E. Shannon, ``Prediction and entropy of printed english,'' \emph{Bell system technical journal}, vol.~30, no.~1, pp. 50--64, 1951.

\bibitem{cox2016syntactically}
D.~Cox, ``Syntactically informed text compression with recurrent neural networks,'' \emph{arXiv preprint arXiv:1608.02893}, 2016.

\bibitem{tian2024synchronous}
Y.~Tian, J.~Ying, Z.~Qin, Y.~Jin, and X.~Tao, ``Synchronous multi-modal semantic communicationsystem with packet-level coding,'' \emph{arXiv preprint arXiv:2408.04535}, 2024.

\bibitem{sullivan2012overview}
G.~J. Sullivan, J.-R. Ohm, W.-J. Han, and T.~Wiegand, ``Overview of the high efficiency video coding (hevc) standard,'' \emph{IEEE Transactions on circuits and systems for video technology}, vol.~22, no.~12, pp. 1649--1668, 2012.

\bibitem{marcellin2000overview}
M.~W. Marcellin, M.~J. Gormish, A.~Bilgin, and M.~P. Boliek, ``An overview of jpeg-2000,'' in \emph{Proceedings DCC 2000. Data compression conference}.\hskip 1em plus 0.5em minus 0.4em\relax IEEE, 2000, pp. 523--541.

\bibitem{schmidhuber1996sequential}
J.~Schmidhuber and S.~Heil, ``Sequential neural text compression,'' \emph{IEEE Transactions on Neural Networks}, vol.~7, no.~1, pp. 142--146, 1996.

\bibitem{mahoney2000fast}
M.~V. Mahoney, ``Fast text compression with neural networks.'' in \emph{FLAIRS}, 2000, pp. 230--234.

\bibitem{goyal2018deepzip}
M.~Goyal, K.~Tatwawadi, S.~Chandak, and I.~Ochoa, ``Deepzip: Lossless data compression using recurrent neural networks,'' \emph{arXiv preprint arXiv:1811.08162}, 2018.

\bibitem{valmeekam2023llmzip}
C.~S.~K. Valmeekam, K.~Narayanan, D.~Kalathil, J.-F. Chamberland, and S.~Shakkottai, ``Llmzip: Lossless text compression using large language models,'' \emph{arXiv preprint arXiv:2306.04050}, 2023.

\bibitem{sutskever2011generating}
I.~Sutskever, J.~Martens, and G.~E. Hinton, ``Generating text with recurrent neural networks,'' in \emph{Proceedings of the 28th international conference on machine learning (ICML-11)}, 2011, pp. 1017--1024.

\bibitem{10.1162/neco.1997.9.8.1735}
\BIBentryALTinterwordspacing
S.~Hochreiter and J.~Schmidhuber, ``Long short-term memory,'' \emph{Neural Comput.}, vol.~9, no.~8, p. 1735–1780, nov 1997. [Online]. Available: \url{https://doi.org/10.1162/neco.1997.9.8.1735}
\BIBentrySTDinterwordspacing

\bibitem{cho2020learning}
K.~Cho, B.~Van~Merrienboer, C.~Gulcehre, D.~Bahdanau, F.~Bougares, H.~Schwenk, and Y.~Bengio, ``Learning phrase representations using rnn encoder-decoder for statistical machine translation. arxiv 2014,'' \emph{arXiv preprint arXiv:1406.1078}, 2020.

\bibitem{pascanu2013difficulty}
R.~Pascanu, ``On the difficulty of training recurrent neural networks,'' \emph{arXiv preprint arXiv:1211.5063}, 2013.

\bibitem{vaswani2017attention}
A.~Vaswani, N.~Shazeer, N.~Parmar, J.~Uszkoreit, L.~Jones, A.~N. Gomez, {\L}.~Kaiser, and I.~Polosukhin, ``Attention is all you need,'' \emph{Advances in neural information processing systems}, vol.~30, 2017.

\bibitem{wu2020deep}
N.~Wu, B.~Green, X.~Ben, and S.~O'Banion, ``Deep transformer models for time series forecasting: The influenza prevalence case,'' \emph{arXiv preprint arXiv:2001.08317}, 2020.

\bibitem{wu2020adversarial}
S.~Wu, X.~Xiao, Q.~Ding, P.~Zhao, Y.~Wei, and J.~Huang, ``Adversarial sparse transformer for time series forecasting,'' \emph{Advances in neural information processing systems}, vol.~33, pp. 17\,105--17\,115, 2020.

\bibitem{dubey2024llama}
A.~Dubey, A.~Jauhri, A.~Pandey, A.~Kadian, A.~Al-Dahle, A.~Letman, A.~Mathur, A.~Schelten, A.~Yang, A.~Fan \emph{et~al.}, ``The llama 3 herd of models,'' \emph{arXiv preprint arXiv:2407.21783}, 2024.

\bibitem{achiam2023gpt}
J.~Achiam, S.~Adler, S.~Agarwal, L.~Ahmad, I.~Akkaya, F.~L. Aleman, D.~Almeida, J.~Altenschmidt, S.~Altman, S.~Anadkat \emph{et~al.}, ``Gpt-4 technical report,'' \emph{arXiv preprint arXiv:2303.08774}, 2023.

\bibitem{anil2023gemini}
R.~Anil, S.~Borgeaud, Y.~Wu, J.-B. Alayrac, J.~Yu, R.~Soricut, J.~Schalkwyk, A.~M. Dai, A.~Hauth, K.~Millican \emph{et~al.}, ``Gemini: A family of highly capable multimodal models,'' \emph{arXiv preprint arXiv:2312.11805}, vol.~1, 2023.

\bibitem{yu2023megabyte}
L.~Yu, D.~Simig, C.~Flaherty, A.~Aghajanyan, L.~Zettlemoyer, and M.~Lewis, ``Megabyte: modeling million-byte sequences with multiscale transformers,'' in \emph{Proceedings of the 37th International Conference on Neural Information Processing Systems}, 2023, pp. 78\,808--78\,823.

\bibitem{wang2024mambabyte}
J.~Wang, T.~Gangavarapu, J.~N. Yan, and A.~M. Rush, ``Mambabyte: Token-free selective state space model,'' \emph{arXiv preprint arXiv:2401.13660}, 2024.

\bibitem{perez2024compressed}
J.~C. P{\'e}rez, A.~Pardo, M.~Soldan, H.~Itani, J.~Leon-Alcazar, and B.~Ghanem, ``Compressed-language models for understanding compressed file formats: a jpeg exploration,'' \emph{arXiv preprint arXiv:2405.17146}, 2024.

\bibitem{wu2024beyond}
S.~Wu, X.~Tan, Z.~Wang, R.~Wang, X.~Li, and M.~Sun, ``Beyond language models: Byte models are digital world simulators,'' \emph{arXiv preprint arXiv:2402.19155}, 2024.

\bibitem{han2024jpeg}
X.~Han, M.~Ghazvininejad, P.~W. Koh, and Y.~Tsvetkov, ``Jpeg-lm: Llms as image generators with canonical codec representations,'' \emph{arXiv preprint arXiv:2408.08459}, 2024.

\bibitem{zlib}
J.~loup Gailly and M.~Adler, ``zlib: A massively spiffy yet delicately unobtrusive compression library,'' \url{https://www.zlib.net/}, 1995, accessed: 2024-12-09.

\bibitem{ziv1977universal}
J.~Ziv and A.~Lempel, ``A universal algorithm for sequential data compression,'' \emph{IEEE Transactions on information theory}, vol.~23, no.~3, pp. 337--343, 1977.

\bibitem{huffman1952method}
D.~A. Huffman, ``A method for the construction of minimum-redundancy codes,'' \emph{Proceedings of the IRE}, vol.~40, no.~9, pp. 1098--1101, 1952.

\bibitem{ma2023intelligent}
H.~Ma, X.~Luo, and D.~Xu, ``Intelligent queue management of open vswitch in multi-tenant data center,'' \emph{Future Generation Computer Systems}, vol. 144, pp. 50--62, 2023.

\bibitem{luo2023clothoid}
X.~Luo, Y.~Feng, and C.~Wang, ``A clothoid curve-based intersection collision warning scheme in internet of vehicles,'' \emph{The Computer Journal}, vol.~66, no.~10, pp. 2447--2461, 2023.

\bibitem{luo2024rm}
X.~Luo, Z.~Li, Z.~Peng, D.~Xu, and Y.~Liu, ``Rm-gen: Conditional diffusion model-based radio map generation for wireless networks,'' in \emph{2024 IFIP Networking Conference (IFIP Networking)}.\hskip 1em plus 0.5em minus 0.4em\relax IEEE, 2024, pp. 543--548.

\bibitem{luo2025denoising}
X.~Luo, Z.~Li, Z.~Peng, M.~Chen, and Y.~Liu, ``Denoising diffusion probabilistic model for radio map estimation in generative wireless networks,'' \emph{IEEE Transactions on Cognitive Communications and Networking}, 2025.

\bibitem{dadkhah2024ciciomt2024}
S.~Dadkhah, E.~Carlos Pinto~Neto, R.~Ferreira, R.~Chukwuka~Molokwu, S.~Sadeghi, and A.~Ghorbani, ``{CICIoMT2024}: Attack vectors in healthcare devices-a multi-protocol dataset for assessing iomt device security,'' \emph{Preprint}, 2024.

\bibitem{zhang2022aoi}
G.~Zhang, C.~Shen, Q.~Shi, B.~Ai, and Z.~Zhong, ``Aoi minimization for wsn data collection with periodic updating scheme,'' \emph{IEEE Transactions on Wireless Communications}, vol.~22, no.~1, pp. 32--46, 2022.

\end{thebibliography}

\end{document}